\documentclass[conference]{IEEEtran}
\IEEEoverridecommandlockouts

\usepackage{cite}
\usepackage{amsmath,amssymb,amsfonts}
\usepackage{graphicx}
\usepackage{textcomp}
\usepackage{xcolor}

\usepackage{url}
\usepackage{booktabs}
\usepackage{tabularx}
\usepackage[caption=false,font=footnotesize]{subfig}
\usepackage{algorithm}
\usepackage[noend]{algpseudocode}
\usepackage{adjustbox}
\usepackage{makecell}

\def\BibTeX{{\rm B\kern-.05em{\sc i\kern-.025em b}\kern-.08em
    T\kern-.1667em\lower.7ex\hbox{E}\kern-.125emX}}

\begin{document}

\title{Capstone: Power-Capped Pipelining for Coarse-Grained Reconfigurable Array Compilers}

\author{
\IEEEauthorblockN{
Sabrina Yarzada and
Christopher Torng
}
\IEEEauthorblockA{
Department of Electrical and Computer Engineering,
University of Southern California, Los Angeles, CA, USA 
}
}

\maketitle


\begin{abstract}
Coarse-grained reconfigurable arrays (CGRAs) combine much of the performance and energy efficiency of ASICs with the flexibility of programmable hardware, making them attractive in power-constrained systems. However, modern CGRA compilers optimize primarily for performance and can create hardware configurations that exceed fixed power budgets. We empirically show that post-place-and-route (post-PnR) pipelining in Cascade, a prior state-of-the-art CGRA compiler, can monotonically increase power as it inserts registers to improve frequency. We present \emph{Capstone}, a power-cap-aware extension of Cascade that integrates a fast, compiler-resident power model with uncertainty-aware controllers to guide bitstream selection. Capstone automatically learns energy costs for compiler-visible abstract events from hierarchical gate-level power supervision and achieves 16.8\% MAPE, approaching a 14.8\% oracle restricted to the same event abstraction, while adding less than 1\% prediction overhead per post-PnR iteration. Capstone I applies a static guardband, Capstone II uses conformal calibration for probabilistic coverage under its calibration assumptions, and Capstone III uses event-level error bounds for a conditional deterministic guarantee. Across eight evaluated vector, matrix, and tensor kernels, all three modes achieve 100\% cap compliance under the evaluated margins. Capstone I, II, and III retain 0.65$\times$, 0.74$\times$, and 0.81$\times$ of uncapped frequency on average while leaving median cap slack of 31.23\%, 22.26\%, and 13.85\%, respectively. Capstone demonstrates that the compiler can proactively optimize the hardware configuration itself for a power constraint before deployment, rather than relying only on reactive run-time power management.
\end{abstract}



\section{Introduction}
\label{sec-intro}

Coarse-grained reconfigurable arrays (CGRAs) have emerged as a useful middle ground between fixed-function ASICs and fine-grained reconfigurable fabrics such as FPGAs~\cite{vasilyev-vision-micro2016, prabhakar-plasticine-isca2017, gobieski-snafu-isca2021, karunaratne-hycube-dac2017, gobieski-riptide-micro2022, torng-uecgra-hpca2021, koul-aha-tecs2023, fan-cgra-objectinf-tvlsi2018, huang-elastic-cgras-fpga2013, jafri-cgra-isqed2013, kim-samsung-cgra-fpt2012}. They retain programmable spatial execution while providing high performance and energy efficiency for applications including machine learning, image processing, sparse tensor computation, and sensor processing. These properties also make CGRAs attractive for systems in which power is a first-order constraint, including thermally limited SoCs, battery-operated devices, and edge accelerators. In these settings, exceeding a power budget can trigger throttling or violate system-level requirements, while leaving excessive unused headroom sacrifices performance that could otherwise be obtained within the same budget.

\begin{figure}[t]
  \centering
  \includegraphics[width=\linewidth]{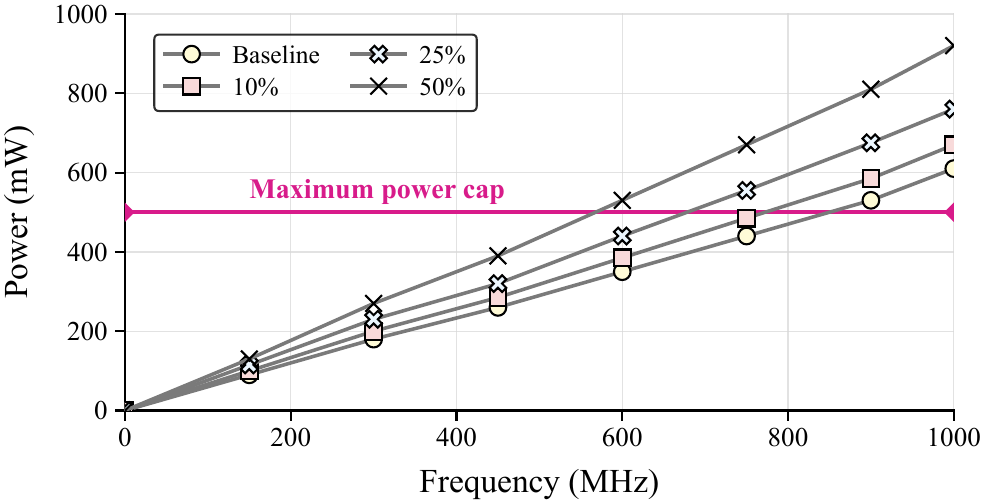}
  \caption{As state-of-the-art CGRA compilers pipeline for higher clock frequencies, they can unknowingly exceed the horizontal power cap. For the vector elementwise addition kernel, projected incremental performance optimizations, such as 10\%, similarly exceed the cap.}
  \label{fig1}
\end{figure}

Yet CGRA compilation remains predominantly performance driven. Scheduling, placement, routing, and pipelining are generally optimized to improve throughput or achievable clock frequency, even though these compiler decisions determine the physical resources exercised by the resulting configuration. Cascade~\cite{melchert-cascade-2024}, for example, performs iterative post-place-and-route (post-PnR) pipelining by inserting configurable registers along routed paths in response to static timing analysis (STA). This approach substantially improves frequency and energy--delay product, but every additional pipelining step can also change register utilization, interconnect activity, routing structure, and control overhead. Our characterization shows that these structural changes can monotonically increase power as Cascade continues to pursue higher frequency, eventually producing configurations that exceed a fixed power cap, as illustrated in Figure~\ref{fig1}.

This behavior exposes a limitation in the traditional separation between compilation and power management. Run-time power-management mechanisms such as DVFS and energy-constrained control~\cite{zoni-energy-constrained-controller-2020,xu-os-2015, chen-energy-2015}, along with run-time power monitoring and modeling~\cite{zoni-runtime-edge-power-monitors-2023,kim-simmani-2019}, are important for responding to workload and operating-condition variation that becomes visible after deployment. However, they operate after a particular hardware configuration has already been selected. If the compiler has inserted unnecessary pipeline registers or selected a structurally expensive routed hardware configuration, reducing its run-time frequency cannot remove those resources or recover the structural overhead they introduced. For reconfigurable architectures, power management therefore has an additional opportunity: act while the compiler can still change the hardware configuration itself.

Doing so requires solving a different power-modeling problem from traditional architectural or run-time power prediction. The backend compiler must evaluate many nearby candidate hardware configurations quickly, yet signoff-quality post-layout power analysis can require hours for each candidate. Our key insight is that CGRA compilers already expose rich structural information describing the architectural datapath, including active tile counts, register insertion, routing density, memory usage, and interconnect utilization. Instead of invoking commercial ASIC signoff tools inside every compiler iteration, the compiler can learn a lightweight model from high-fidelity supervision and evaluate that model using its own intermediate representations. We use the term \emph{compiler-resident power model} to refer to a model that runs inside the compiler, consumes compiler-visible features, and is fast enough to guide optimization decisions directly. We train this model automatically from matched compiler event vectors and hierarchical gate-level power reports, jointly learning a nonnegative mapping between the two representations and the corresponding per-event energy coefficients. The learned hierarchy is then folded into a fast sum-of-products predictor over compiler-visible events. We also compare the model against a non-deployable oracle restricted to the same event vocabulary, allowing us to distinguish limitations of model training from limitations of the compiler abstraction itself. By embedding such a model into the post-PnR pipelining loop, the compiler can treat power as a first-class optimization objective alongside timing and shift from maximizing frequency alone to maximizing \emph{safe} frequency under a power cap.


We present Capstone\footnote{The software and supporting artifacts for this work are publicly available at \url{https://github.com/usc-acorn/yarzada-capstone-micro2026}}, a power-cap-aware extension of Cascade~\cite{melchert-cascade-2024} that integrates this fast, compiler-resident power model calibrated to commercial ASIC signoff data with a user-tunable controller that guides post-PnR pipelining toward cap-compliant points. During each iteration, Capstone predicts candidate power and selects one or a small set of PnR configurations such that at least one satisfies a user-specified power cap under the selected uncertainty model. Capstone I uses a static guardband and requires no calibration data. Capstone II uses conformal calibration to construct a learned safety envelope with a probabilistic guarantee under standard calibration assumptions. Capstone III uses user-specified event-level error bounds to reason over a bounded family of possible models, providing a conditional hard guarantee when the true error lies within the specified bounds. These modes span early-stage design exploration, calibrated pre-silicon evaluation, and conservative deployment with explicit uncertainty bounds.

Capstone deliberately targets post-PnR pipelining because Cascade~\cite{melchert-cascade-2024} creates a clear and repeated power and frequency tradeoff as it changes the hardware configuration for higher performance. Even after scheduling and routing are fixed, Capstone can avoid power-unsafe structural over-pipelining and select better cap-aware operating points. More broadly, Capstone provides a general methodology for power-aware structural compilation that requires only compiler-visible activity features and high-fidelity power supervision, and could therefore be extended to scheduling, placement, routing, memory organization, and other compiler-controlled decisions in reconfigurable and spatial architectures.

Our primary contributions are as follows:
\begin{itemize}

\item We develop a hierarchical, machine-learned energy model trained using gate-level power estimation supervision. The model jointly learns (i)~compiler-visible energy events, (ii)~their per-event energy costs, and (iii)~a mapping to fine-grained RTL power instances, bridging architectural and physical representations to enable fast compile-time power estimation without invoking signoff tools inside the compiler loop.
\item We design power-aware compiler controllers that explicitly account for model uncertainty through static guardbands, conformal safety envelopes, and bounded-error planning, allowing the compiler to return cap-aware candidate bitstreams while maximizing achievable frequency.
\item We integrate Capstone into Cascade~\cite{melchert-cascade-2024} and evaluate it across dense and sparse CGRA kernels using a commercial ASIC flow, demonstrating cap-compliant bitstream selection with low compile-time overhead while preserving useful performance under fixed power constraints.
\end{itemize}


\begin{figure}[t]
  \centering

  \begin{minipage}[c]{0.6\columnwidth}
    \centering
    \includegraphics[width=0.95\columnwidth]{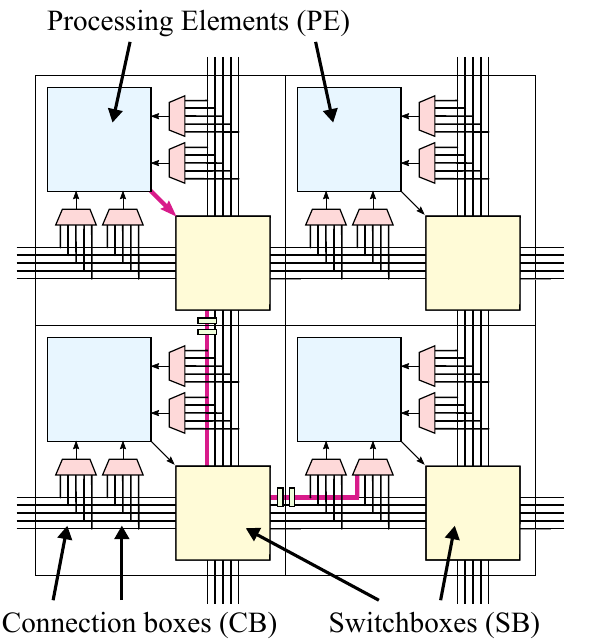}
  \end{minipage}\hfill%
  \begin{minipage}[c]{0.4\columnwidth}
    \caption{CGRA architecture including PE tiles and interconnect components
    (SBs, CBs). The magenta route highlights a representative routed critical
    path that post-PnR pipelining breaks by inserting green registers along
    the interconnect. Input and output tracks along the north and west
    perimeters are connected to memory banks.}
    \label{fig2}
  \end{minipage}

\end{figure}

To our knowledge, we are the first to propose an automated, data-driven method to train an energy and power model for abstract events rooted directly in the software stack, in this case the backend compiler of a CGRA, that after training repeatably approaches the accuracy of an oracle that is restricted to the same set of abstract events. We are also the first to demonstrate that an automated compiler-level optimization can proactively tune the hardware itself for a power and energy constraint with a formalized data-driven confidence level at configuration time, rather than reactively relying on architectural power management techniques deployed only at run time.


\section{Background and Related Work}
\label{sec-background-related}
We first establish why CGRA backends expose a configuration-time power optimization opportunity, then review power modeling and prior energy-aware CGRA compilation, and finally contrast Capstone's uncertainty-aware structural optimization with run-time power management.

\paragraph{CGRAs and compiler-controlled structural optimization} Coarse-grained reconfigurable arrays (CGRAs)~\cite{vasilyev-vision-micro2016, prabhakar-plasticine-isca2017, gobieski-snafu-isca2021, karunaratne-hycube-dac2017, gobieski-riptide-micro2022, torng-uecgra-hpca2021, koul-aha-tecs2023, fan-cgra-objectinf-tvlsi2018, huang-elastic-cgras-fpga2013, jafri-cgra-isqed2013, kim-samsung-cgra-fpt2012} are programmable spatial accelerators that combine architectural flexibility with high performance and energy efficiency. As illustrated in Figure~\ref{fig2}, a representative CGRA consists of processing-element and memory tiles connected by a programmable interconnect, with applications mapped through scheduling, placement, routing, and bitstream generation. These compiler decisions determine which compute, memory, routing, and optional pipeline resources are exercised, so different configurations of the same application can have different performance and power characteristics.

\begin{figure}[t]
  \centering
  \includegraphics[width=\linewidth]{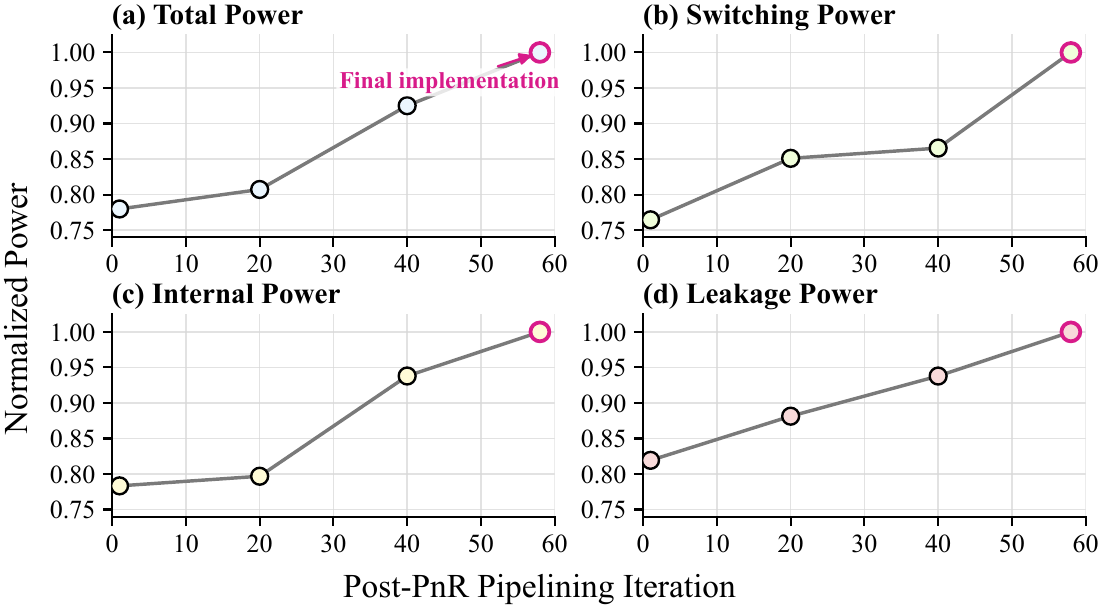}
  \caption{Normalized power versus post-PnR pipelining iteration for the VecAdd kernel under Cascade~\cite{melchert-cascade-2024}. Panels show (a) total, (b) switching, (c) internal, and (d) leakage power, each normalized to the final implementation. Power generally increases as Cascade inserts pipeline registers to improve timing, with the final maximally pipelined implementation highlighted in magenta.}
  \label{fig3}
\end{figure}

Cascade~\cite{melchert-cascade-2024} provides a concrete example of this structural optimization. Its post-PnR pipelining pass iteratively inserts configurable registers along critical paths using STA feedback. Although this improves frequency and energy--delay product, the compiler does not observe the corresponding power cost during the loop. As shown in Figure~\ref{fig3}, increasing pipeline depth can increase power, creating a discrete structural power / performance search space. This makes post-PnR pipelining an opportunity to consider power while the compiler can still change the implementation itself.

\paragraph{Power models from higher-level abstractions}
Power estimation spans several abstraction levels, trading fidelity for evaluation cost~\cite{coburn-power-2005,jianlei-power-2015, kim-strober-2016}. Post-PnR gate-level analysis with extracted parasitics provides high fidelity but is too slow for repeated use inside an iterative compiler loop. Architectural and learning-based~\cite{zhang-grannite-2020, zhou-primal-2019, nasser-power-2021, lin-hl-pow-2020, lee-power-model-2018, kumar-cpu-power-2023} models instead estimate power from higher-level information, including microarchitectural features, HLS representations, RTL activity, and netlist structure~\cite{li-mcpat-taco2013,zhou-primal-2019, zhang-grannite-2020,lin-hl-pow-2020,lee-power-model-2018, kumar-cpu-power-2023}. Simmani~\cite{kim-simmani-2019}, for example, automatically selects RTL signals for run-time power modeling, while PowerProbe~\cite{zoni-powerprobe-2018} derives run-time models through automatic RTL instrumentation. NeuroMeter~\cite{tang-neurometer-2021} instead provides fast microarchitecture-level power, area, and timing models for machine-learning accelerators.

Capstone targets a different abstraction boundary. That is, abstract events already defined by the backend compiler's software representation. It uses hierarchical gate-level power supervision to automatically learn both nonnegative per-event energy costs and their mapping to fine-grained physical power contributors. The resulting \emph{compiler-resident power model} runs directly from compiler-visible state and is fast enough to guide inner-loop optimization. Because this event representation necessarily omits fine-grained switching and physical effects, we also compare Capstone against an oracle restricted to the same abstract event vocabulary, allowing us to distinguish limitations of model training from limitations of the compiler abstraction itself.

\paragraph{Energy-aware CGRA compilation}
Prior work has already demonstrated that energy can influence CGRA mapping and configuration decisions. Peng et al. propose a CGRA energy model and optimize loop mapping to reduce battery charge loss~ \cite{peng-battery-aware-cgra-aspdac2015}. Gu et al. integrate multi-$V_{\mathrm{DD}}$ assignment with CGRA loop mapping to reduce energy without degrading performance~ \cite{gu-multivdd-cgra-aspdac2017}. Nalesh et al. study energy-aware synthesis of application datapaths on a CGRA~ \cite{nalesh-energy-aware-cgra-integration2017}, while Kojima et al. jointly optimize configurable pipeline structure and body bias under performance constraints~\cite{kojima-body-bias-cgra-ieice2018}. These works establish that CGRA mappings and configurations can be optimized for energy rather than performance alone.

Capstone addresses a complementary problem. Rather than relying on a predefined energy model or technology-specific optimization objective, it automatically learns the power and energy costs of abstract compiler events from fine-grained physical supervision and then uses that learned model to enforce a user-specified power constraint during iterative backend structural optimization. In particular, Capstone associates this decision with an explicit uncertainty model rather than assuming that the compiler-level power estimate is exact.

\paragraph{Uncertainty-aware optimization and run-time power management}
Robust optimization under bounded uncertainty, chance-constrained optimization, and conformal prediction provide complementary mechanisms for decision-making under model uncertainty. Set- and budget-based robust optimization protects against bounded perturbations~\cite{bertsimas-sim-robustness-2004, srivastava-robust-opt-2007}, while chance-constrained methods enforce probabilistic feasibility under uncertain parameters~\cite{ nemirovski-shapiro-chance-2006,zhang-chance-constraints-2011, duan-chance-constraints-2018}. Hardware guardbanding similarly uses explicit safety margins to tolerate PVT and aging variation~\cite{ rahimi-guardband-2013,rahimi-guardband-2014}. Conformal prediction constructs calibrated prediction sets or bounds from held-out data~\cite{shafer-vovk-conformal-2008,lu-conformal-pred-calib-2023, yeh-conformal-pred-calib-2024}. Capstone adapts these ideas to compiler-level structural optimization: Capstone I uses a static guardband, Capstone II uses calibration residuals to construct a conformal upper bound, and Capstone III uses bounded event-level error to construct a robust upper envelope.

Run-time power management has been studied extensively across CPUs, GPUs, accelerators, and SoCs~\cite{zoni-runtime-edge-power-monitors-2023, zoni-energy-constrained-controller-2020,xu-os-2015, chen-energy-2015}. DVFS, throttling, and sensor-driven control are essential for reacting to workload and operating-condition variation after deployment. Energy-aware execution and reconfiguration have similarly been explored for FPGAs and heterogeneous accelerators~
\cite{tann-runtime-config-2016,xu-dynamic-2017,liu-runtime-2013, nabavinejad-batchsizer-2021,nabavinejad-batching-dvfs-2022, he-hybrid-gpufpga-ics2021}. These mechanisms can modify an operating point or execution strategy, but they generally act after or around implementation selection. A run-time controller can lower the voltage or frequency of a deployed bitstream, for example, but cannot remove pipeline registers or routing overhead already introduced during compilation. Capstone instead applies an uncertainty-aware power constraint while the backend compiler can still alter the implementation, leaving run-time control as a complementary mechanism for residual deployment variation.

In summary, prior work demonstrates energy-aware CGRA mapping, higher-level power modeling, uncertainty-aware optimization, and run-time power management. Capstone connects these areas by automatically learning a physically supervised power model for abstract events rooted in the compiler software stack. It uses this model with explicit data-driven uncertainty to guide structural optimization under a power constraint at configuration time.
\newcommand{\E}{\mathbb{E}}
\newcommand{\R}{\mathbb{R}}
\newcommand{\ptpx}{\textsc{PrimeTime\,PX}}

\section{Capstone: Power-Aware Compilation Under Model Uncertainty}
\label{sec-capstone}

The preceding sections motivate two requirements for power-aware structural compilation. First, the backend compiler needs a physically grounded power estimate that can be evaluated fast enough to guide its optimization loop. Second, because any compiler-level model necessarily omits some physical information, the compiler must account explicitly for prediction uncertainty when deciding whether a configuration satisfies a power cap. Capstone addresses these requirements with two corresponding components: (i)~an automatically trained \emph{compiler-resident power model} and (ii)~an uncertainty-aware controller for configuration-time bitstream selection.

We use \emph{configuration time} to refer to the pre-deployment backend compilation stage in which the CGRA implementation and bitstream are formed and selected. In Cascade~\cite{melchert-cascade-2024}, the relevant structural degree of freedom is post-PnR pipelining. Successive candidates contain different inserted pipeline registers and therefore represent different physical implementations. Capstone operates while these structural choices can still be changed, before run time.

The most direct way to obtain accurate power for each candidate would be to export every intermediate PnR configuration, construct the corresponding hardware netlist, and invoke signoff-quality power analysis. This is the procedure used to generate the characterization in Figure~\ref{fig3}, but each signoff invocation can require hours and is therefore unsuitable for an inner compiler loop. Capstone instead performs expensive physical supervision offline and distills it into a compiler-resident power model that can be evaluated from compiler-visible state during post-PnR pipelining.

\begin{table}[t]
\centering
\caption{Capstone energy model notation.}
\label{tbl1}

\small
\setlength{\tabcolsep}{2.5pt}
\renewcommand{\arraystretch}{1.03}

\begin{tabularx}{\columnwidth}{
@{}
l
>{\raggedright\arraybackslash}X
@{\hspace{0.8em}}
l
>{\raggedright\arraybackslash}X
@{}
}
\toprule
\textbf{Symbol} & \textbf{Description} &
\textbf{Symbol} & \textbf{Description} \\
\midrule

$E,R$ &
\makecell[l]{Number of events,\\\ptpx{} rows} &
$e,r,k$ &
\makecell[l]{Event, row,\\dataset indices} \\

$X^{(k)}$ &
\makecell[l]{Compiler event-count\\/activity vector} &
$X_e^{(k)}$ &
\makecell[l]{Count of event $e$\\in dataset $k$} \\

$y^{(k)}$ &
\makecell[l]{\ptpx{} row-\\power vector} &
$\hat{y}^{(k)}$ &
\makecell[l]{Predicted row-\\power vector} \\

$\hat{P}^{(k)}$ &
\makecell[l]{Predicted total\\power} &
$\mathbf{1}$ &
\makecell[l]{All-ones\\vector} \\

$W,W_{re}$ &
\makecell[l]{Event-to-row\\allocation weights} &
$\alpha,\alpha_e$ &
\makecell[l]{Per-event energy\\coefficients} \\

$\beta_e$ &
\makecell[l]{Effective compile-\\time event cost} &
$P_{\mathrm{leak}}$ &
\makecell[l]{Leakage-power\\term} \\

$\lambda_W,\lambda_\alpha$ &
\makecell[l]{Regularization\\weights} &
$b$ &
\makecell[l]{Operating-point\\bin} \\

$V,f,T$ &
\makecell[l]{Voltage, frequency,\\temperature} &
$\rho(f)$ &
\makecell[l]{Frequency-scaling\\factor} \\

$\mathrm{MAPE}$ &
\makecell[l]{Mean absolute\\percentage error} &
$R^2$ &
\makecell[l]{Coefficient of\\determination} \\

\bottomrule
\end{tabularx}
\end{table}
At each iteration, Capstone uses this model to estimate candidate power, converts the estimate into a mode-specific upper bound, and uses that bound to determine whether the candidate is admissible under a user-specified power cap. The controller attempts to return a conservative \emph{anchor} configuration satisfying the selected uncertainty condition and may optionally expose additional speculative candidates when the deployment can perform further validation or provide a run-time fallback. Capstone I uses a static engineering guardband, Capstone II derives its margin from held-out calibration data, and Capstone III reasons over bounded event-level model error. Thus, all three modes share the same compiler-resident model and candidate search while differing in the assumptions used to bound prediction uncertainty.

\subsection{Energy Model Learning with Gate-Level Supervision}
\label{sec-hier-energy-model}

\begin{figure}[t]
  \centering
  \includegraphics[width=1.0\linewidth]{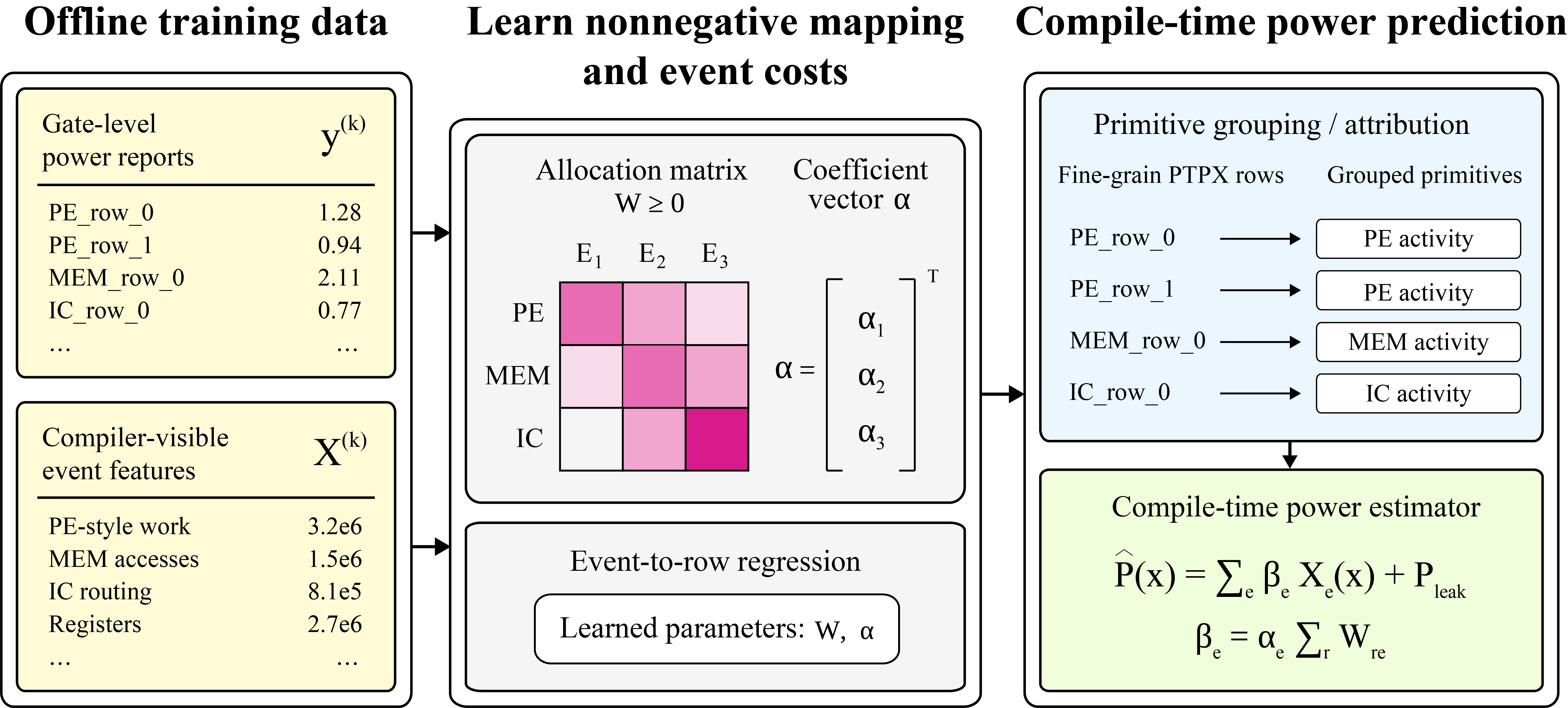}
  \caption{Hierarchical energy model learning. Capstone learns a nonnegative mapping $W$ and per-event coefficients $\alpha$ that align compiler events to gate-level rows, discover energy events from per-row power, then perform compile-time prediction via a sum-of-products over compiler-visible features.}
  \label{fig5}
\end{figure}

We develop a data-driven, hierarchical methodology for learning compiler-visible energy models in which signoff-quality gate-level power analysis supervises the learning of per-event energy costs and their mapping to compiler-level activity features. In our implementation, we use Synopsys \ptpx{} (PTPX) as the signoff power estimation tool. The objective is to retain the \emph{physical grounding} of gate-level measurements while delivering a fast \emph{compile-time} estimator suitable for inner-loop optimization. Figure~\ref{fig5} summarizes the training flow. For a fixed architecture, workload realization, mapped configuration, and operating point, the compiler and the signoff tool observe the same execution at different levels of abstraction. The compiler observes architectural events such as active PE operations, memory accesses, routing activity, and inserted registers, while PTPX reports gate-level instance power using the post-layout netlist, parasitics, and activity information available in the signoff flow. Capstone learns a nonnegative mapping from compiler-visible events to PTPX power-report rows together with the corresponding per-event energy costs, retaining physical grounding while producing a compiler-resident power model that can be evaluated without rerunning signoff analysis. We summarize notation in Table~\ref{tbl1}.

This use of signoff power analysis is confined to offline model supervision and signoff tools are not invoked inside the compiler optimization loop. Their post-layout capacitance information and gate-level hierarchy are nevertheless valuable because they expose the physical cost of routing, registers, memory structures, and control logic that purely analytical compiler models can miss. The resulting compiler-resident model is approximate because some information is compressed or unavailable at compile time, including fine-grained toggle rates, duty cycles, data-dependent switching, local parasitic variation, and differences between model-fitting and deployment activity. Capstone accounts for this residual error through the uncertainty-aware controllers in Section~\ref{sec-controllers}, rather than treating the learned prediction as exact.

This methodology is not specific to PTPX or to the CGRA studied in this work. It requires only (i)~compiler-visible activity features obtained by exercising the datapath with representative input data and (ii)~a signoff-quality power estimation tool that provides a per-instance or hierarchical power breakdown. By learning the mapping from compiler event space to signoff power-report rows over multiple datasets, Capstone replaces manual feature-to-report mapping with a learned nonnegative allocation and an automated regression procedure that can be updated as additional labeled examples become available.

\paragraph{Hierarchical model formulation}
Capstone learns the relationship between what the compiler can observe and what the signoff power tool reports. For each compiled configuration, the compiler provides a set of activity measurements, such as the number of active PE operations, memory accesses, inserted registers, and routed interconnect events. For the same configuration, the signoff power tool reports how much power is consumed by physical structures in the implemented design, such as PE tiles, memories, registers, and interconnect blocks. Capstone uses many matched examples of these two views to learn how compiler-visible events contribute to physical power. Figure ~\ref{fig5} illustrates this relationship.

We represent one matched example by an index $k$. Let $E$ denote the number of compiler-visible event types and let $R$ denote the number of rows in the signoff power report. For example, one event type may count register activity and one signoff row may report the power consumed by a register hierarchy. For example $k$, let

\[
X^{(k)} \in \mathbb{R}_{\geq 0}^{E}
\]

be the vector of compiler-visible event activity. Its element $X_e^{(k)}$ is the activity associated with compiler event $e$ for that configuration. The compiler initially observes structural event counts. Capstone scales these counts by the corresponding event rate when predicting power, so $X_e^{(k)}$ represents the effective activity used by the power model. For the same example, let

\[
y^{(k)} \in \mathbb{R}_{\geq 0}^{R}
\]

be the vector of power values reported by the signoff tool. Its element $y_r^{(k)}$ is the measured power of signoff report row $r$. In our implementation, these rows are provided by PTPX and correspond to entries in its gate-level power hierarchy.

Capstone learns two quantities that connect these representations. First, each compiler event $e$ is assigned a nonnegative energy coefficient $\alpha_e$. The vector

\[
\alpha \in \mathbb{R}_{\geq 0}^{E}
\]

collects these coefficients and represents the average energy cost of the compiler-visible events. Thus, $\alpha_e$ captures \emph{how much} event $e$ costs on average.

Second, Capstone learns a nonnegative allocation matrix

\[
W \in \mathbb{R}_{\geq 0}^{R\times E}.
\]

Each column of $W$ describes the \emph{allocation pattern} of one compiler event across the signoff power report rows. Each element $W_{re}$ indicates how strongly compiler event $e$ contributes to signoff power-report row $r$. Thus, while $\alpha_e$ captures \emph{how much} event $e$ costs, $W_{re}$ captures \emph{where} that event's energy appears in the physical design. In this sense, $W$ learns which physical power report rows are associated with each compiler-visible event.

Using these learned quantities, Capstone predicts the full signoff power breakdown for example $k$ as

\begin{equation}
\hat{y}^{(k)}
=
W\,\mathrm{diag}(\alpha)\,X^{(k)},
\label{eq-predictor}
\end{equation}

where $\hat{y}^{(k)}\in\mathbb{R}_{\geq0}^{R}$ is the predicted row-by-row power vector and $\mathrm{diag}(\alpha)$ is the diagonal matrix containing the learned event-energy coefficients. Intuitively, Equation~\ref{eq-predictor} first assigns an energy cost to each compiler-visible event and then distributes that contribution across the physical power-report rows using $W$. Figure~\ref{fig5} illustrates this process where $\alpha$ sets the energy magnitude of each event, while the corresponding column of $W$ distributes that contribution among the PE, memory, and interconnect rows.

The hierarchical prediction is useful during model fitting because it preserves the correspondence between compiler events and physical structures. The compiler itself, however, does not need this full row-by-row breakdown during every optimization iteration. We therefore collapse the learned hierarchy into one effective coefficient per compiler event.

For event $e$, define

\begin{equation}
\beta_e
\triangleq
\alpha_e
\sum_{r=1}^{R} W_{re},
\label{eq-effective-coeff}
\end{equation}

where $\beta_e$ is the effective compile-time energy cost of event $e$ after summing its contributions across all signoff power-report rows. Thus, $\alpha_e$ is the per-event energy scale used in the hierarchical model during fitting, while $\beta_e$ is the corresponding folded per-event coefficient used by the compiler-resident model. The compiler-resident power prediction then becomes

\begin{equation}
\hat{P}^{(k)}
=
\sum_{e=1}^{E}
\beta_e X_e^{(k)}
+
P_{\mathrm{leak}},
\label{eq-compile}
\end{equation}

where $\hat{P}^{(k)}$ is the predicted total power for example $k$ and $P_{\mathrm{leak}}$ is the modeled leakage-power contribution. Equation~\ref{eq-compile} is the model evaluated inside Cascade's post-PnR optimization loop. After offline fitting, the compiler only needs the compiler-visible activity $X^{(k)}$, the learned effective coefficients $\beta$, and the leakage term for evaluation, as shown in Figure~\ref{fig5}. This reduction preserves the information learned from hierarchical physical supervision while making prediction inexpensive enough for repeated inner-loop use.

\paragraph{Model fitting}
The fitting objective uses the signoff hierarchy to constrain the event-to-row mapping while encouraging sparse and stable assignments:

\begin{equation}
\min_{W,\alpha\ge 0}\;
\sum_{k}
\big\|y^{(k)}-\hat{y}^{(k)}\big\|_2^2
+
\lambda_W\|W\|_1
+
\lambda_\alpha\|\alpha\|_2^2.
\label{eq-energy-model-objective}
\end{equation}

The nonnegativity constraints preserve physical interpretability because compiler events cannot contribute negative power. The $\ell_1$ penalty encourages each event to map to a small number of meaningful signoff power report rows, while the $\ell_2$ penalty stabilizes the learned event costs. This structure is intentionally more constrained than a generic black-box predictor. Although a larger unconstrained model could potentially fit total power more aggressively, Capstone preserves nonnegative event costs, row-level physical attribution, and the event-level structure needed to construct the uncertainty bounds used by Capstone III in Section~\ref{sec-controllers}.

We solve Equation~\ref{eq-energy-model-objective} using alternating nonnegative updates. Holding $\alpha$ fixed, each column of $W$ is updated using nonnegative least squares with an $\ell_1$ penalty. Holding $W$ fixed, $\alpha$ is updated using nonnegative ridge regression. After each update of $W$, we normalize its columns and transfer the corresponding scale into $\alpha$, eliminating the ambiguity in which $W$ and $\alpha$ can be rescaled in opposite directions without changing the prediction. The model is initialized by connecting each compiler event uniformly to the available signoff power-report rows and fitting an initial $\alpha$ to total power. Hyperparameters $(\lambda_W,\lambda_\alpha)$ are selected using validation, and fitting stops when validation error no longer improves.

\paragraph{Fitting coverage and abstraction error}
Capstone fits the model using multiple \emph{activity datasets}. Each dataset contains a compiler event vector and a matched signoff power-report vector for a particular kernel, input/activity realization, mapped configuration, and operating point. The datasets may therefore span different input distributions, schedules, pipelining depths, and routing variants. Including intermediate pipelining iterations and alternative routes broadens coverage of the compiler-visible event space, whereas fitting on a narrow set of examples can increase error when deployment candidates exercise unseen combinations of registers, routing tracks, memory paths, or compute resources.

\begin{figure}[t]
  \centering
  \includegraphics[width=\linewidth]{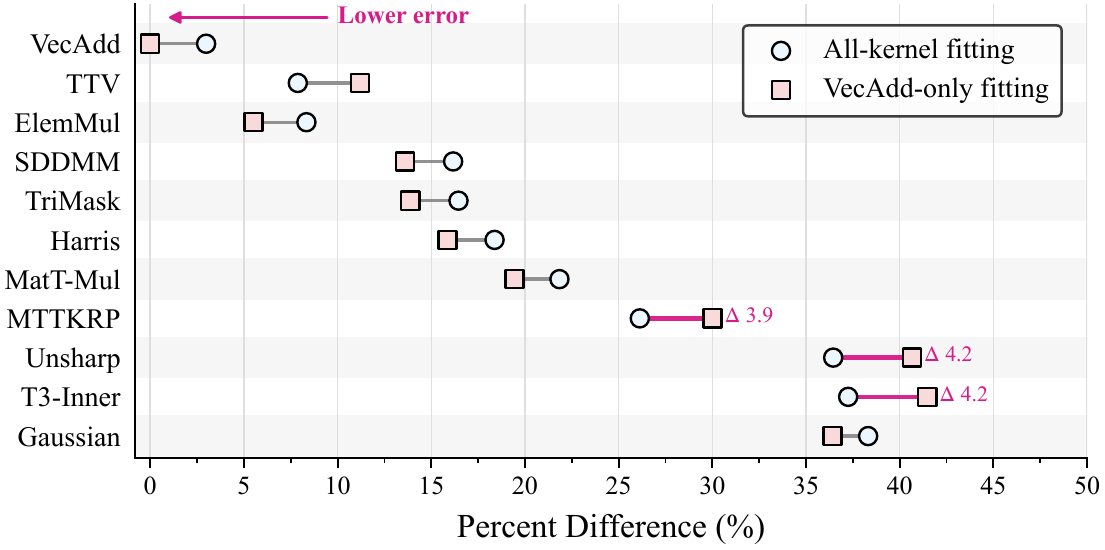}
\caption{Per-kernel percent difference between signoff PTPX and Capstone estimates under all-kernel and VecAdd-only model fitting. Lower values indicate better agreement with PTPX. Magenta annotations highlight the largest error increases under VecAdd-only fitting.}
  \label{fig4}
\end{figure}

Figure~\ref{fig4} illustrates this effect across kernels. A model fit using examples from the broader kernel set generally agrees more closely with PTPX than one fit using VecAdd alone, particularly for larger kernels whose activity patterns differ substantially from VecAdd. Although broader fitting reduces their error, some of these kernels still have higher error because they exercise more behavior that is not fully captured by the compiler-visible event vocabulary. This motivates fitting over diverse compiler-visible activity patterns while also accounting for residual abstraction error. Capstone therefore records the feature ranges observed during model fitting and controller calibration. If a deployment candidate falls outside these ranges, the controller can fall back to a coarser residual group, apply a larger margin, request additional calibration data, or require additional signoff or silicon-calibrated measurements (Section~\ref{sec-controllers}).

Even with broad fitting coverage, some contributors to physical power are not explicitly available as compiler-visible prediction features. These include bit-level switching, data-dependent toggle rates, glitching, duty cycles, detailed physical placement effects, local parasitics, and variation in event bitwidth. As such, the learned parameters represent expectations over hidden physical and activity factors under the model-fitting distribution. In this sense, $\alpha_e$ captures the expected energy associated with event $e$, while $W_{re}$ captures the expected fraction of that event's contribution attributed to power report row $r$.

The compiler-resident model is consequently most accurate when deployment activity resembles the examples used for fitting and may degrade when the workload, routing structure, activity distribution, or operating point shifts substantially. This residual abstraction error is expected rather than treated as a failure of the model. It motivates both the uncertainty-aware controller design in Section~\ref{sec-controllers} and the held-out and cross-kernel evaluation in Section~\ref{sec-eval}.

\paragraph{Operating-point dependence}
Capstone explicitly accounts for operating-point dependence across both the energy model and the compiler-level controllers. The learned coefficients represent event energy for a supported voltage, process corner, and temperature regime. Frequency is handled primarily through event rate, as increasing frequency causes the same energy per event to contribute greater dynamic power. Accordingly, Capstone converts compiler-visible event counts to prediction-time activity through scaling by frequency and initiation interval.


\begin{figure}[t]
  \centering
  \includegraphics[width=\linewidth]{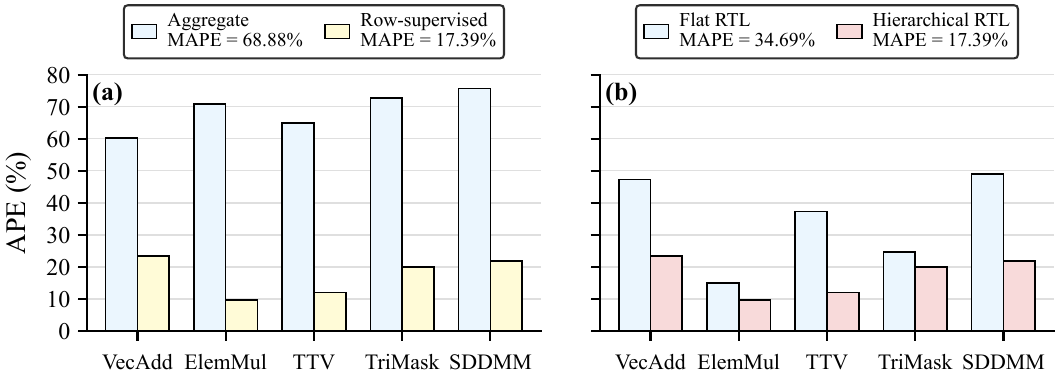}
  \caption{Per-kernel APE for Capstone power models under (a) aggregate versus row-supervised calibration and (b) flat versus hierarchical RTL modeling. Lower APE is better. Legends report MAPE across the evaluated kernels.}
  \label{fig6}
\end{figure}
Since voltage, process corner, and temperature can change both event energy and leakage power, a deployment flow using Capstone can fit separate models for supported operating point bins. Alternatively, it can augment the controller with explicit margins derived from corner sweeps, technology characterization, or post-silicon measurements. In Section~\ref{sec-controllers}, these bins are denoted $b=(V,\mathrm{corner},T)$. Any remaining frequency-dependent model error is handled separately through the controller scaling factor $\rho(f)$. This separation keeps the learned event energy interpretation clear while allowing the controller to account for operating conditions not exactly represented by the model.

\paragraph{Hierarchical versus aggregate supervision}
To isolate the value of hierarchical signoff supervision, Capstone also supports a simpler \emph{scalar aggregate} model. This baseline omits $W$ and fits event coefficients directly to total reference power,

\begin{equation}
\hat{P}^{(k)}
=
\sum_e \alpha_e X_e^{(k)}
+
P_{\mathrm{leak}},
\label{eq-scalar-baseline}
\end{equation}

using nonnegative regression with a small ridge penalty. The scalar model remains fast and interpretable, but total-power supervision alone does not constrain how individual compiler events correspond to physical power contributors.

The hierarchical model instead uses the full signoff power-report vector $y^{(k)}$ and jointly learns the event costs and event-to-row allocation. Row-level supervision provides more constraints than fitting only total power, reduces ambiguity among correlated compiler events, and produces a physical decomposition that can be inspected at the level of tiles, memories, registers, and interconnect structures. Figure~\ref{fig6}(a) shows the practical benefit of this additional supervision. Across the evaluated kernels, the scalar aggregate model has 68.88\% MAPE, while the row-supervised model reduces MAPE to 17.39\%. This result motivates using the available physical hierarchy rather than learning event costs from total power alone.

The signoff reports used for fitting may themselves expose different amounts of hierarchy. Rich hierarchical reports group power rows by architectural structures such as compute tiles, memories, interconnect, registers, and control logic, which makes the learned columns of $W$ more informative and can improve model accuracy. Flatter reports remain compatible with the methodology because Capstone requires only a set of reported power contributors, but they provide less information about where individual event costs arise physically.

Figure~\ref{fig6}(b) evaluates this distinction. Using a flatter report increases MAPE to 34.69\%, compared with 17.39\% when the available hierarchical report structure is retained. Accordingly, Capstone uses hierarchical signoff power reports when available while remaining compatible with flatter report organizations.

\paragraph{Offline versus compiler-loop cost}
\begin{table}[t]
\centering
\caption{Capstone model setup and compile-time cost.}
\label{tbl-model-setup-cost}

\small
\setlength{\tabcolsep}{2.2pt}
\renewcommand{\arraystretch}{1.05}

\resizebox{\columnwidth}{!}{%
\begin{tabular}{@{}lccc@{}}
\toprule
\textbf{Component} &
\textbf{Stage} &
\textbf{Cost} &
\textbf{Amortized over} \\
\midrule

PTPX labels &
Offline &
$\sim$36 h/design &
Design \\

Feature extraction &
Offline/compile &
$<1$ s/candidate &
Candidate \\

Alt. regression &
Offline/op. bin &
Min./arch./bin &
Architecture/bin \\

Conformal calib. &
Offline/group/bin &
Sec./group/bin &
Group/bin \\

Capstone prediction &
Compiler loop &
$<1\%$ of iteration &
Candidate \\

Signoff power &
Offline &
Hours/candidate &
Not in loop \\

\bottomrule
\end{tabular}%
}
\end{table}

Table~\ref{tbl-model-setup-cost} separates Capstone's offline model construction cost from its online prediction overhead. The dominant offline expense is generating labeled signoff examples, which are produced by PTPX in our implementation. The alternating nonnegative regression is also performed offline and is amortized over compiler invocations for a given architecture and operating-point bin.

During Cascade's post-PnR pipelining loop, Capstone neither reruns PTPX nor refits Equation~\ref{eq-energy-model-objective}. It only extracts the compiler-visible event activity and evaluates the folded sum-of-products predictor in Equation~\ref{eq-compile}. This separation between expensive offline physical supervision and lightweight online inference is what allows the model to be compiler-resident. Capstone retains the physical information learned from signoff while adding less than 1\% prediction overhead per post-PnR iteration in our evaluation.

\subsection{Compiler-Level Power-Aware Controllers}
\label{sec-controllers}

\begin{figure}[t]
  \centering
  \includegraphics[width=\linewidth]{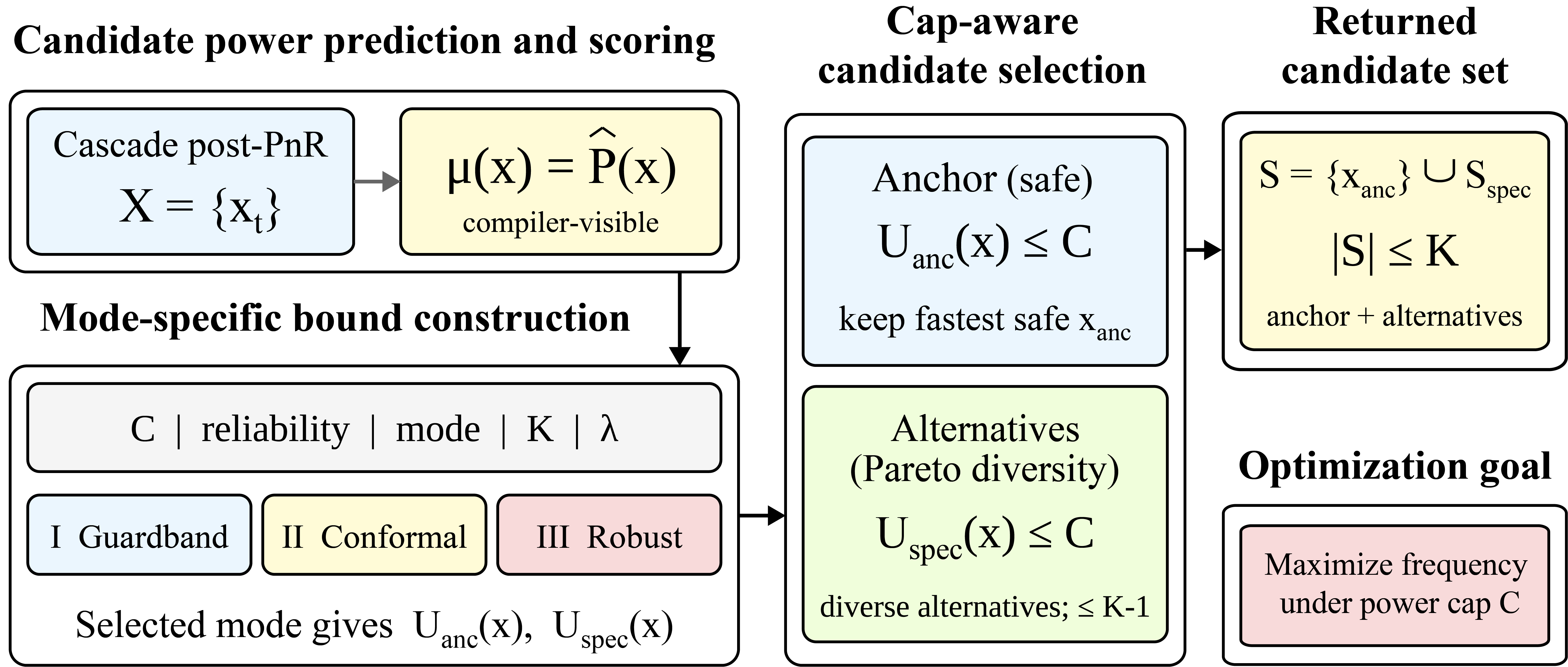}
\caption{Capstone optimization and controller framework. Cascade post-PnR candidates are scored with compiler-visible power estimates and converted into mode-specific anchor and speculative bounds using Capstone I (guardband), II (conformal), or III (robust). The highest-frequency candidate satisfying the conservative anchor bound is retained as the safe anchor. When $K>1$, additional candidates satisfying the speculative bound may be selected using Pareto diversity to expose alternative frequency--slack tradeoffs. Capstone returns the anchor and at most $K-1$ alternatives while maximizing achievable frequency under the power cap $C$.}
  \label{fig8}
\end{figure}

\begin{table}[t]
\centering
\caption{Capstone controller notation.}
\label{tbl2}

\small
\setlength{\tabcolsep}{2.5pt}
\renewcommand{\arraystretch}{1.03}

\begin{tabularx}{\columnwidth}{
@{}
l
>{\raggedright\arraybackslash}X
@{\hspace{0.8em}}
l
>{\raggedright\arraybackslash}X
@{}
}
\toprule
\textbf{Symbol} & \textbf{Description} &
\textbf{Symbol} & \textbf{Description} \\
\midrule

$x,\mathcal{X}$ &
\makecell[l]{Compiler config.,\\candidate sequence} &
$f(x),\Delta f$ &
\makecell[l]{Achievable freq.,\\minimum step} \\

$C$ &
Power cap &
$S,x_{\mathrm{anc}}$ &
\makecell[l]{Returned set,\\anchor bitstream} \\

$S_{\mathrm{spec}},\lambda$ &
\makecell[l]{Speculative set,\\diversity weight} &
$\mu(x),P^{\mathrm{true}}(x)$ &
\makecell[l]{Predicted power,\\true power} \\

$P_i^{\mathrm{ref}}$ &
\makecell[l]{High-fidelity\\reference power} &
$U_\star(x)$ &
\makecell[l]{Power bound for\\$\star\in\{\mathrm{anc},\mathrm{spec}\}$} \\

$U_{\mathrm{gb}},U_{\mathrm{conf}}$ &
\makecell[l]{Guardband bound,\\conformal bound} &
$U_{\mathrm{rob}}$ &
\makecell[l]{Bounded-error\\robust bound} \\

$\gamma_\star$ &
\makecell[l]{Static guardband\\parameter} &
$\tau_{\mathrm{anc}},\alpha_\star$ &
\makecell[l]{Confidence target,\\miscoverage level} \\

$D_{\mathrm{cal}}$ &
\makecell[l]{Held-out\\calibration set} &
$g,b$ &
\makecell[l]{Residual group,\\operating-point bin} \\

$s_i,r_i$ &
\makecell[l]{Raw residual,\\normalized residual} &
$q_\alpha,n_{g,b},k_\alpha$ &
\makecell[l]{Quantile, group\\size, quantile index} \\

$\rho(f)$ &
\makecell[l]{Frequency-scaling\\factor} &
$e,X_e(x)$ &
\makecell[l]{Event index,\\event count/activity} \\

$\beta_e,\epsilon_e$ &
\makecell[l]{Event cost,\\event error bound} &
$\epsilon_{e,\cdot}$ &
\makecell[l]{Fit/activity and\\PVT/OOD terms} \\

$P_{\mathrm{leak}},\epsilon_{\mathrm{leak}}$ &
\makecell[l]{Leakage power,\\leakage error bound} &
$K$ &
Returned set size limit \\

\bottomrule
\end{tabularx}
\end{table}

Capstone uses the compiler-resident power model from Section~\ref{sec-hier-energy-model} to guide bitstream selection during Cascade's post-PnR pipelining loop. Figure~\ref{fig8} summarizes the resulting optimization and controller framework. During each pipelining iteration, Cascade produces a candidate configuration $x$ with a routed graph, achievable frequency $f(x)$, initiation interval, and compiler-visible event activity. Capstone evaluates each candidate using the compiler-resident power prediction $\mu(x)=\hat{P}(x)$ and determines whether that candidate is admissible under a user-specified power cap $C$.

Because $\mu(x)$ is an approximate prediction rather than signoff power, the controller does not compare it directly against the cap. Instead, it converts the point estimate into an uncertainty-aware upper bound on true power. Capstone provides three controller modes, summarized in Table~\ref{tbl-controller-bounds}, that share the same candidate search but differ in how they construct this bound. Capstone~I applies a static guardband, Capstone~II derives a probabilistic upper bound from held-out calibration residuals, and Capstone~III constructs a deterministic robust upper bound from explicit event-level error intervals. This separation allows the same compiler-resident model and post-PnR search to support different assumptions about model uncertainty and deployment information.

The controller attempts to return at least one conservative \emph{anchor} bitstream satisfying the selected mode's cap-compliance condition. If the user requests more than one deployment option, the controller may additionally expose \emph{speculative} candidates that offer higher frequency or less unused power headroom. We denote the maximum number of returned bitstreams by $K$ and summarize the remaining controller notation in Table~\ref{tbl2}.

The anchor is the bitstream that carries the controller's mode-specific safety claim. If $K=1$, Capstone returns only this anchor. If $K>1$, speculative candidates may use a less conservative bound to expose additional performance. These candidates do not inherit the anchor's guarantee unless they satisfy the same bound. Capstone does not select among the $K$ speculative candidates at run time. A deployment that uses a speculative candidate must provide its own selection and validation mechanism, such as signoff checking, on-chip power sensing with fallback, throttling, or reverting to the anchor.

\begin{table}[t]
\centering
\caption{Capstone controller modes and upper bounds.}
\label{tbl-controller-bounds}

\small
\setlength{\tabcolsep}{3pt}
\renewcommand{\arraystretch}{1.08}

\begin{tabularx}{\columnwidth}{
@{}
>{\raggedright\arraybackslash}p{0.36\columnwidth}
>{\raggedright\arraybackslash}X
@{}
}
\toprule
\textbf{Controller and Upper Bound} &
\textbf{Interpretation} \\
\midrule

\makecell[tl]{
\textbf{Capstone I}\\[1pt]
$U_{\mathrm{gb}}(x)=(1+\gamma)\mu(x)$
}
&
Static guardband with no calibration requirement or statistical guarantee.
Safety requires $\gamma$ to cover model error, PVT variation, activity
mismatch, and deployment shift.
\\[4pt]

\makecell[tl]{
\textbf{Capstone II}\\[1pt]
$U_{\mathrm{conf}}(x)=\mu(x)$\\
$+\rho(f(x))q_{\alpha}(g(x),b(x))$
}
&
Conformal envelope providing $1-\alpha$ coverage under exchangeability
within group $g$ and bin $b$. Coarser groups are used when calibration
data is limited.
\\[4pt]

\makecell[tl]{
\textbf{Capstone III}\\[1pt]
$U_{\mathrm{rob}}(x)=$\\
$\sum_e(\beta_e+\epsilon_e)X_e(x)$\\
$+P_{\mathrm{leak}}+\epsilon_{\mathrm{leak}}$
}
&
Bounded-error robust controller providing a deterministic anchor guarantee
when true event costs lie within the specified error intervals. Returns at
most $K$ bitstreams.
\\

\bottomrule
\end{tabularx}
\end{table}
\paragraph{Shared candidate selection}
Algorithm~\ref{alg-capstone-controller} gives the selection procedure shared by all three controller modes. Candidates are considered in increasing frequency. Whenever a higher-frequency candidate satisfies the conservative anchor bound $U_{\mathrm{anc}}(x)$, it becomes the new anchor. If $K>1$, additional candidates may also be retained using a speculative bound $U_{\mathrm{spec}}(x)$.

Speculative candidates are retained only if they satisfy their corresponding bound and are not dominated by another candidate. A candidate is discarded when another retained candidate provides both higher frequency and at least as much safety slack, or when the two are nearly identical in compiler-visible features. This pruning keeps the returned set bounded by $K$ while preserving the anchor and therefore does not weaken the anchor's mode-specific guarantee.

\paragraph{Calibration and operating-point coverage}
Capstone distinguishes \emph{model fitting} from \emph{controller calibration}. Model fitting learns the event costs and event-to-row allocation described in Section~\ref{sec-hier-energy-model}. Controller calibration estimates how much additional margin is required to use the resulting compiler-resident power prediction during bitstream selection.

\begin{algorithm}[t]
\caption{Generic Capstone Controller}
\label{alg-capstone-controller}
\begin{algorithmic}[1]
\Require Candidate sequence $\mathcal{X}$; power cap $C$; output limit $K$; upper bounds $U_{\mathrm{anc}}(\cdot)$ and $U_{\mathrm{spec}}(\cdot)$; minimum frequency step $\Delta f$; diversity weight $\lambda$
\Ensure Up to $K$ bitstreams, including one conservative anchor
\State Initialize:
$S_{\mathrm{spec}}\gets \emptyset$, $x_{\mathrm{anc}}\gets \bot$, $f_{\mathrm{prev}}\gets 0$

\For{each pipelined configuration $x\in\mathcal{X}$ in increasing frequency}
    \If{$f(x)<f_{\mathrm{prev}}+\Delta f$}
        \State \textbf{continue}
    \EndIf
    \If{$U_{\mathrm{anc}}(x)\le C$}
        \State $x_{\mathrm{anc}}\gets x$
    \EndIf
    \If{$K>1$ and $U_{\mathrm{spec}}(x)\le C$}
        \State Insert $x$ into $S_{\mathrm{spec}}$ by frequency, slack, and diversity
        \State Pareto-prune and retain up to $K-1$ candidates
    \EndIf
    \State $f_{\mathrm{prev}}\gets f(x)$
\EndFor
\If{$x_{\mathrm{anc}}=\bot$}
    \State \Return failure or request a lower frequency target, larger cap, or different mapping
\EndIf
\State \Return $\{x_{\mathrm{anc}}\}\cup S_{\mathrm{spec}}$
\end{algorithmic}
\end{algorithm}
When calibration is used, the calibration set is held out from the examples used to fit the energy model and contains configurations with both compiler predictions and high-fidelity reference power:

\begin{equation}
D_{\mathrm{cal}}
=
\{(x_i,P_i^{\mathrm{ref}},\mu_i,g_i,f_i,b_i)\}_{i=1}^{n}.
\label{eq:controller-calibration-set}
\end{equation}

Here, $x_i$ is calibration configuration $i$,
$P_i^{\mathrm{ref}}$ is its high-fidelity reference power from signoff
or a silicon-calibrated measurement,
$\mu_i=\hat{P}(x_i)$ is the compiler-resident prediction,
$g_i$ is an optional residual-group label,
$f_i$ is the candidate frequency, and
$b_i$ is the operating-point bin.

For each calibration kernel and activity realization, the calibration set can include both the final pipelining point and intermediate candidates selected by frequency, pipeline depth, or feature-space quantiles. Final points exercise high register and interconnect activity, while intermediate points better represent configurations that may be selected under tighter power caps. The group label $g_i$ can identify a kernel family, workload class, or structural regime. When a group contains too few samples for stable calibration, Capstone falls back to a coarser group or the global residual distribution.

The calibration set also defines the deployment regime for which a calibrated bound is intended to apply. The controller is most reliable when deployment kernels, input and activity distributions, routing patterns, and operating points resemble those represented during calibration. As such, Capstone checks whether a candidate's compiler-visible features remain within the calibrated feature ranges. If a candidate falls outside this region, or if the deployment workload differs substantially from the calibration set, the controller can use a more conservative global margin, widen its guardband, request additional calibration examples, or require additional signoff or
silicon measurements.

Capstone handles voltage, process, and temperature variation using operating-point bins. Each bin $b=(V,\mathrm{corner},T)$ corresponds to a supported voltage, process corner, and temperature range. A deployment flow can fit a separate model and residual table for each supported bin or add an explicit PVT margin derived from signoff corner sweeps, technology characterization, or post-silicon measurements. Frequency is handled primarily through the event-rate scaling in the compiler-resident power model. If validation shows that residual prediction error also grows with frequency, the controller can scale its uncertainty margin using a factor $\rho(f)$.

\paragraph{Capstone I: Static Guardband Controller}
Capstone~I provides the simplest way to convert the compiler-resident prediction into a conservative power estimate by not requiring a calibration set. It applies a multiplicative guardband
$\gamma$:

\begin{equation}
U_{\mathrm{gb}}(x)
=
(1+\gamma)\mu(x).
\label{eq:guardband-bound}
\end{equation}

A configuration is accepted when
$U_{\mathrm{gb}}(x)\le C$. The anchor uses a conservative guardband
$\gamma_{\mathrm{anc}}$, while optional speculative candidates may use
a smaller guardband $\gamma_{\mathrm{spec}}$, with
$\gamma_{\mathrm{anc}}\ge\gamma_{\mathrm{spec}}$, when the user wants
to expose higher-frequency candidates for later validation.

Capstone~I does not provide a statistical confidence guarantee. Its cap-compliance claim is conditional on the chosen guardband covering the true relative underestimation error in deployment. If $\gamma_{\mathrm{anc}}$ upper-bounds that error, then an anchor satisfying $U_{\mathrm{gb}}(x)\le C$ also satisfies the power cap.

The guardband must therefore account for sources of underestimation not explicitly captured by the compiler-resident model, including model error, activity mismatch, PVT variation, frequency-scaling error, and deployment shift. When validation data are available, a practical choice is

\begin{equation}
\gamma_{\mathrm{anc}}
\ge
Q_{1-\alpha}
\left(
\max\left(
0,
\frac{P_i^{\mathrm{ref}}-\mu_i}{\mu_i}
\right)
\right)
+
\gamma_{\mathrm{PVT}}
+
\gamma_{\mathrm{OOD}},
\label{eq:guardband-selection}
\end{equation}

where the first term is a high quantile of observed relative underestimation, $\gamma_{\mathrm{PVT}}$ covers expected operating-condition variation, and $\gamma_{\mathrm{OOD}}$ covers expected deployment shift. If no validation data are available, $\gamma$ must instead be chosen from conservative engineering judgment or technology-level characterization. Thus, Capstone~I is therefore best viewed as a guardband-based engineering mechanism rather than a calibrated confidence mechanism.

Because unmodeled deployment shifts can still exceed a chosen guardband, Capstone~I can be paired with run-time fallback mechanisms such as power-meter-triggered throttling, DVFS reduction, or frequency capping~\cite{reddi-arch-throttling-2009,
kim-arch-throttling-2006,
aragon-arch-throttling-2006,
lee-arch-throttling-2006}. Wider guardbands increase conservatism and
leave more unused power headroom, while narrower guardbands preserve more performance but place greater reliance on later validation or run-time correction. A full exploration of these fallback mechanisms is orthogonal to this work.

\paragraph{Capstone II: Conformal-Envelope Controller}
When suitable held-out calibration data are available, Capstone~II replaces the manually selected scalar guardband with a one-sided residual margin derived using conformal calibration~
\cite{lu-conformal-pred-calib-2023,
yeh-conformal-pred-calib-2024,
gopakumar-conformal-pred-calib-2024,
cohen-conformal-pred-calib-2025,
cocheteux-conformal-pred-calib-2025}.
The relevant failure mode for power capping is model \emph{underestimation}, since overestimation only causes the compiler to be more conservative. Thus, Capstone forms the nonnegative calibration residual

\begin{equation}
s_i
=
\max(0,P_i^{\mathrm{ref}}-\mu_i).
\label{eq:conformal-residual}
\end{equation}

If the residual magnitude depends on frequency, Capstone normalizes each calibration residual by $\rho(f_i)$ and restores this scaling when constructing the deployment bound. A conservative choice is

\[
\rho(f)=\max(1,f/f_{\mathrm{ref}})
\]

when residual error grows approximately with frequency.

For a miscoverage level $\alpha$, residual group $g$, and operating-point bin $b$, let

\[
r_i=\frac{s_i}{\rho(f_i)}
\]

be the normalized residuals from calibration examples satisfying
$g_i=g$ and $b_i=b$, and let
$r_{(1)}\le\cdots\le r_{(n_{g,b})}$ denote those residuals in sorted
order. Capstone uses the finite-sample conformal index

\begin{equation}
k_\alpha
=
\left\lceil
(n_{g,b}+1)(1-\alpha)
\right\rceil,
\qquad
q_\alpha(g,b)=r_{(k_\alpha)}.
\label{eq:conformal-order-statistic}
\end{equation}

If $k_\alpha>n_{g,b}$, the requested finite-sample conformal quantile cannot be supported by that group. Capstone then falls back to a coarser group, the global residual distribution, additional calibration data, or a separately justified engineering margin.

For anchor and speculative candidates, the resulting upper bound is

\begin{equation}
\begin{aligned}
U_{\mathrm{conf}}^{\star}(x)
&=
\mu(x)
+
\rho(f(x))
q_{\alpha_{\star}}(g(x),b(x)),\\
&\hspace{1.5em}
\star\in\{\mathrm{anc},\mathrm{spec}\}.
\end{aligned}
\label{eq:conformal-bound}
\end{equation}

For the anchor, let
$\tau_{\mathrm{anc}}\in(0,1)$ denote the requested coverage level.
Capstone sets

\begin{equation}
\alpha_{\mathrm{anc}}
=
1-\tau_{\mathrm{anc}},
\label{eq:anchor-alpha}
\end{equation}

and uses $q_{\alpha_{\mathrm{anc}}}$ to construct $U_{\mathrm{conf}}^{\mathrm{anc}}$. Under exchangeability between calibration and deployment examples within the selected residual group and operating-point bin,

\begin{equation}
\Pr\left(
P^{\mathrm{true}}(x)
\le
U_{\mathrm{conf}}^{\mathrm{anc}}(x)
\right)
\ge
\tau_{\mathrm{anc}}.
\label{eq:conformal-coverage}
\end{equation}

Therefore, if the returned anchor satisfies

\[
U_{\mathrm{conf}}^{\mathrm{anc}}(x_{\mathrm{anc}})
\le C,
\]

then it satisfies the power cap with coverage at least
$\tau_{\mathrm{anc}}$ under the stated calibration assumptions.

The grouping choice trades margin tightness against sample size and the validity of the calibration assumption. Fine-grained groups, such as a kernel family or routing-structure class, can produce smaller margins when enough calibration examples are available, but small groups may not support the desired finite-sample quantile. As a result, Capstone uses coarser or global residuals when necessary. Speculative candidates may use a larger $\alpha_{\mathrm{spec}}$ to admit higher-frequency designs, but they do not carry the anchor's coverage claim unless the same target is applied or they are subsequently validated through signoff or run-time sensing.


\begin{figure}[t]
  \centering
  \includegraphics[width=\linewidth]{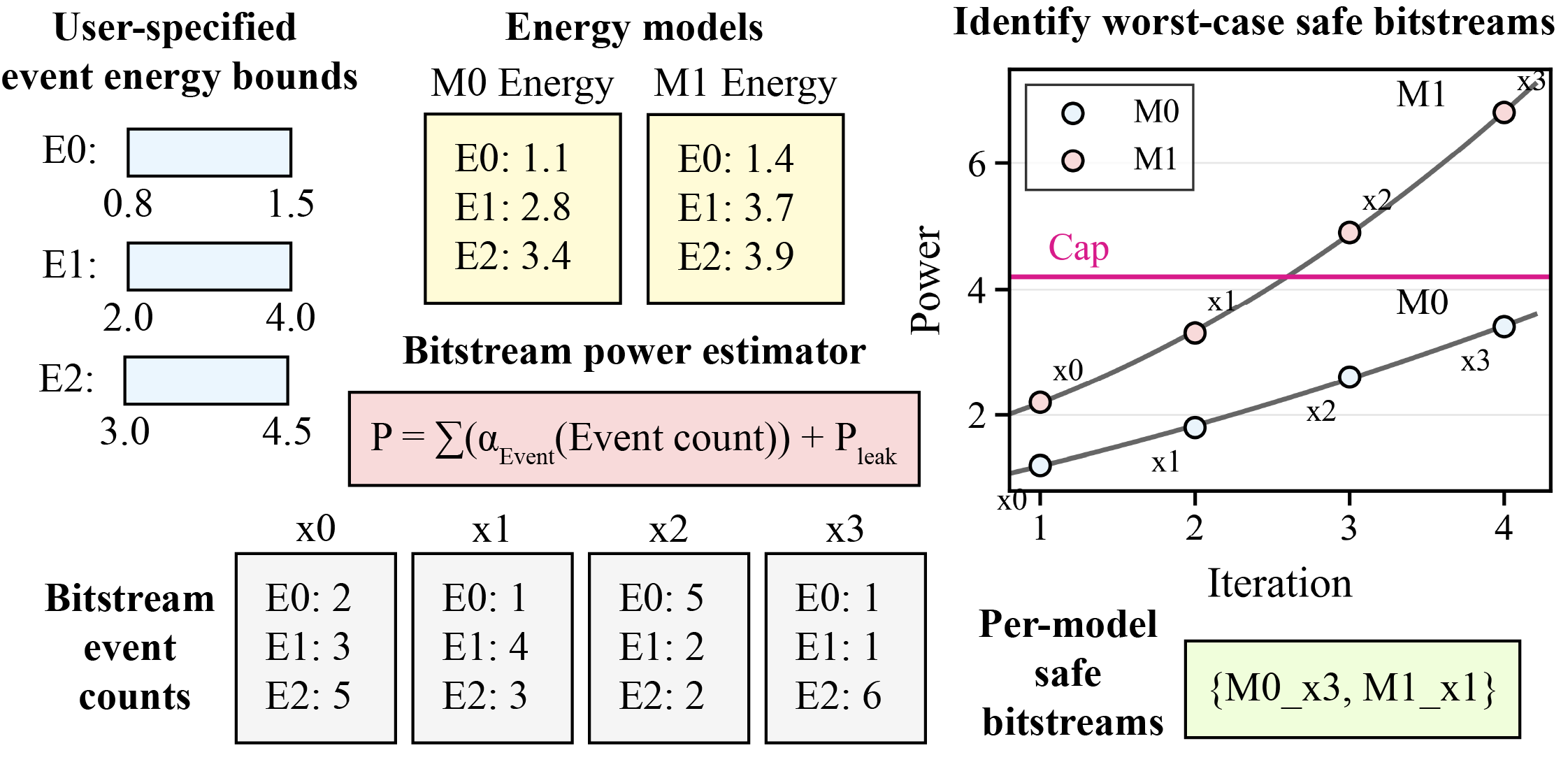}
\caption{Illustrative example of bounded-error power modeling in Capstone~III. $E0$--$E2$ are compiler-visible event types, with blue intervals showing their allowed energy-coefficient ranges in normalized energy units. $M0$ and $M1$ are example energy models within these bounds, and $x0$--$x3$ are candidate bitstreams with the event counts shown below. $M_j\_x_i$ denotes the power predicted for bitstream $x_i$ under model $M_j$. The plot shows predicted power in normalized units, and the magenta line is the user-specified cap. Here, $x3$ is feasible under $M0$, while only up to $x1$ is feasible under $M1$, showing how event-energy uncertainty can change which bitstream is considered safe.}
  \label{fig9}
\end{figure}
\paragraph{Capstone III: Bounded-Error Robust Controller}
Capstone~III replaces the distributional calibration assumption of Capstone~II with explicit bounds on the learned event costs. Figure~\ref{fig9} illustrates this view. Starting from the deployed compiler-resident model, each event coefficient may vary within a specified interval, defining a bounded family of possible power models. Capstone~III selects an anchor that remains below the power cap for every model in this family.

Recall that the compiler-resident power model is

\begin{equation}
\mu(x)
=
\sum_e \beta_e X_e(x)
+
P_{\mathrm{leak}},
\label{eq:robust-base-model}
\end{equation}

where $\beta_e$ is the folded per-event energy coefficient from Section~\ref{sec-hier-energy-model}, $X_e(x)\ge0$ is the compiler-visible activity for event $e$, and $P_{\mathrm{leak}}$ is the modeled leakage contribution. Suppose the corresponding true deployment quantities satisfy

\begin{equation}
\beta_e^{\mathrm{true}}
\in
[\beta_e-\epsilon_e,\beta_e+\epsilon_e],
\qquad
P_{\mathrm{leak}}^{\mathrm{true}}
\le
P_{\mathrm{leak}}
+
\epsilon_{\mathrm{leak}},
\label{eq:event-error-interval}
\end{equation}

where $\epsilon_e$ bounds the possible error in event coefficient
$\beta_e$ and $\epsilon_{\mathrm{leak}}$ bounds leakage error.

Rather than enumerating every model inside these intervals, Capstone~III evaluates the worst-case upper envelope directly:

\begin{equation}
U_{\mathrm{rob}}(x)
=
\sum_e
(\beta_e+\epsilon_e)X_e(x)
+
P_{\mathrm{leak}}
+
\epsilon_{\mathrm{leak}}.
\label{eq:robust-bound}
\end{equation}

Because every event activity $X_e(x)$ is nonnegative, $U_{\mathrm{rob}}(x)$ upper-bounds the predicted power of every model inside the specified uncertainty set. Therefore, if

\[
U_{\mathrm{rob}}(x)\le C,
\]

then $P^{\mathrm{true}}(x)\le C$ for every event-cost assignment contained in those bounds. Capstone~III provides a deterministic cap-compliance guarantee, conditional on the specified intervals containing the true deployment error.

The event-level uncertainty bounds combine expected sources of deployment error:

\begin{equation}
\epsilon_e
=
\epsilon_{e,\mathrm{fit}}
+
\epsilon_{e,\mathrm{act}}
+
\epsilon_{e,\mathrm{PVT}}
+
\epsilon_{e,\mathrm{OOD}}.
\label{eq:event-error-decomposition}
\end{equation}

Here,
$\epsilon_{e,\mathrm{fit}}$ captures model-fitting uncertainty estimated
from cross-validation, held-out signoff error, or coefficient variation
across folds;
$\epsilon_{e,\mathrm{act}}$ captures activity and workload mismatch;
$\epsilon_{e,\mathrm{PVT}}$ captures voltage, process, temperature, and
DVFS variation; and
$\epsilon_{e,\mathrm{OOD}}$ captures expected deployment shift.
When per-event attribution is unavailable or
fine-grained, the same construction can be applied at the
primitive-bucket level, such as PE, memory, register, switchbox, and
routing-mux events, or using a relative bound
$\epsilon_e=\eta_r\beta_e$ for all events in primitive bucket $r$.

To clarify, Capstone~III does not return one bitstream for every possible model. It first selects the highest-frequency candidate satisfying $U_{\mathrm{rob}}(x)\le C$ as the robust anchor. If $K>1$, Capstone first removes Pareto-dominated speculative candidates. If more than $K-1$ speculative candidates remain, it sorts them by frequency and uniformly samples $K-1$ candidates across the remaining Pareto frontier. Along with the robust anchor, this limits the output to at most $K$ candidates. Since the robust anchor is always retained, pruning speculative candidates does not weaken its guarantee. If every returned bitstream must satisfy the same deterministic condition, Capstone applies $U_{\mathrm{rob}}(x)\le C$ to all candidates before pruning.

If independent worst-case error bounds are overly conservative, Capstone can instead use a budgeted uncertainty model in which only a limited number of event groups are allowed to reach their worst-case error simultaneously. This reduces pessimism while preserving a deterministic guarantee under the stated budgeted-error assumption. Thus, Capstone~III replaces Capstone~II's distributional calibration assumption with an explicit bounded-error assumption.

\paragraph{Summary of controller modes}
The three Capstone modes differ in the information available to bound compiler-model error. Capstone~I requires only a user-selected guardband and provides a conditional engineering margin without a statistical guarantee. Capstone~II uses held-out calibration residuals to provide probabilistic coverage under its calibration assumptions. Capstone~III uses explicit event-level error intervals to provide a deterministic guarantee when the true deployment model lies within the specified uncertainty set. Across all three modes, Capstone uses the same compiler-resident power model and post-PnR candidate search to maximize achievable frequency under a user-specified power cap while making the corresponding uncertainty assumptions explicit.
\begin{table}[t]
\centering
\caption{Evaluation benchmarks.}
\label{tbl-benchmarks}

\small
\setlength{\tabcolsep}{2.0pt}
\renewcommand{\arraystretch}{1.04}

\begin{tabularx}{\columnwidth}{
@{}
l
>{\raggedright\arraybackslash}X
@{\hspace{0.9em}}
l
>{\raggedright\arraybackslash}X
@{}
}
\toprule
\textbf{Kernel} & \textbf{Operation} &
\textbf{Kernel} & \textbf{Operation} \\
\midrule

ElemMul &
\makecell[l]{Element-wise matrix\\multiplication} &
SDDMM &
\makecell[l]{Sampled dense-dense\\multiplication} \\

MatT-Mul &
\makecell[l]{Matrix-transpose\\multiplication} &
MTTKRP &
\makecell[l]{Tensor Khatri-Rao\\product} \\

T3-Inner &
\makecell[l]{3D tensor inner\\product} &
Harris &
\makecell[l]{Harris corner\\detection} \\

VecAdd &
Vector element-wise addition &
TTV &
Tensor-times-vector \\

TriMask &
Triangular masking &
InnerProd &
Inner product \\

Gaussian &
Gaussian filtering &
Unsharp &
Unsharp masking \\

\bottomrule
\end{tabularx}
\end{table}
\section{Evaluation}
\label{sec-eval}
Our evaluation answers five questions. First, how accurately does Capstone's compiler-resident power model predict signoff power, and how much of the remaining error comes from the compiler-visible event abstraction? Second, what overhead does this prediction add to Cascade's post-PnR pipelining loop? Third, how do the three Capstone controllers trade off frequency, unused power headroom, and signoff-verified cap compliance? Fourth, how sensitive are the controllers to calibration coverage and bounded-error assumptions? Fifth, how does configuration-time structural selection compare with power models and run-time mechanisms that adjust only the operating point of an already selected implementation?

\subsection{Experimental Setup}
\begin{table}[t]
\centering
\caption{Power estimator baselines evaluated on the same Capstone candidate set.}
\label{tbl-model-baselines}
\small
\setlength{\tabcolsep}{2.8pt}
\renewcommand{\arraystretch}{1.02}

\begin{tabular*}{\columnwidth}{@{\extracolsep{\fill}}lccc@{}}
\toprule
\textbf{Estimator} &
\textbf{Supervision} &
\textbf{Features} &
\textbf{MAPE} \\
\midrule

Analytical primitive &
Manual prim. costs &
Prim. counts &
16.24\% \\

Scalar agg. NNLS &
Total PTPX power &
Event counts &
68.88\% \\

Capstone row-sup. &
PTPX row power &
Events + hier. &
17.39\% \\

Oracle &
PTPX ground truth &
Same abstraction &
14.80\% \\

\bottomrule
\end{tabular*}
\end{table}

Our experiments target a \(32\times16\) CGRA fabric generated by Garnet~\cite{StanfordAHA_garnet} that we then implement in Intel 16\,nm technology. It consists of PE, MEM, and I/O tiles connected through a programmable routing interconnect. Interconnect segments support optional pipeline registers inserted by Cascade during iterative post-PnR pipelining. We hold the fabric parameters fixed across all experiments and evaluate dense, sparse, and image-processing kernels from the Cascade flow, summarized in Table~\ref{tbl-benchmarks}. Dense kernels are generated using Halide~\cite{ragan-halide-2013} and Clockwork~\cite{huff-clockwork-2021} and represented as CoreIR~\cite{daly-coreir-2018} dataflow graphs, while sparse kernels are generated by Custard and translated to the SAM~\cite{hsu-sparse-sam-asplos2023} dataflow representation.

We use a Cascade-derived~\cite{melchert-cascade-2024} compiler flow for scheduling, placement, routing, and post-PnR pipelining, augmented with Capstone's compiler-resident power model and controller modes. All experiments use the same Cadence and Synopsys ASIC toolchain. Cascade's STA reports achievable frequency at each post-PnR pipelining iteration, while Capstone evaluates power directly from compiler-visible event activity. No signoff power analysis or gate-level activity extraction is invoked inside the compiler optimization loop.

For the high-fidelity physical reference required during offline model fitting and evaluation, we use commercial post-PnR signoff power analysis with PTPX. For each routed application graph, we generate RTL and perform synthesis, place-and-route, parasitic extraction, and gate-level power analysis using switching activity from compiled testbenches to obtain hierarchical power breakdowns.


\begin{figure}[t]
  \centering
  \includegraphics[width=\linewidth]{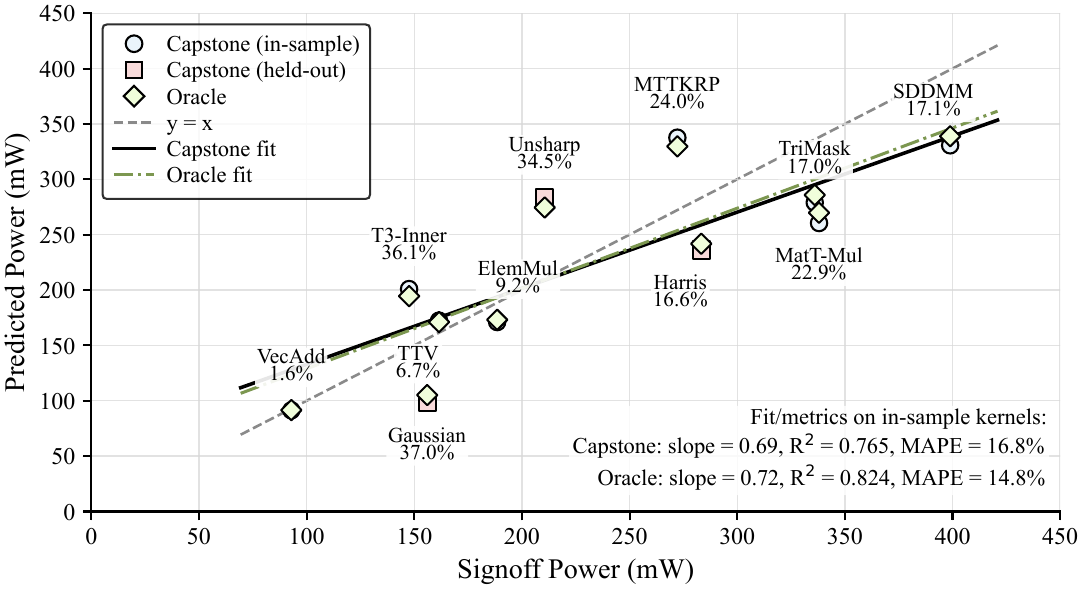}
  \caption{Capstone model accuracy against signoff power at 100 MHz in Intel 16\,nm technology. Labels report Capstone per-kernel APE relative to PTPX, corresponding to vertical deviation from the \(y=x\) reference. Oracle and fitted trends characterize residual model bias and abstraction error.}
  \label{fig10}
\end{figure}

Because no physical CGRA platform with the required sensor infrastructure is available, signoff power serves as the reference for model fitting and evaluation. This leaves a gap relative to silicon deployment, where voltage droop, temperature gradients, process variation, aging, sensor noise, and workload-dependent activity can shift realized power. Capstone treats these effects as sources of uncertainty handled by its controller modes. Our results support the paper's pre-silicon claims about compiler-level power modeling and cap-aware bitstream selection under a consistent signoff reference, while silicon-calibrated validation remains future work.

Unless otherwise noted, model accuracy and controller results are evaluated at matched operating points. This reflects the intended use of the model in Section~\ref{sec-hier-energy-model}, where event-energy coefficients are fit for a supported voltage, process corner, and temperature bin. Cross-operating-point deployment can instead use separate models for \(b=(V,\mathrm{corner},T)\) bins or explicit controller margins derived from signoff corner sweeps, technology characterization, or silicon measurements.


\begin{figure}[t]
  \centering

  \subfloat[Event-to-PTPX mapping.\label{fig12-a}]{%
    \begin{minipage}[t]{0.58\columnwidth}
      \vspace{0pt}
      \centering
      \scriptsize
      \setlength{\tabcolsep}{3pt}
      \renewcommand{\arraystretch}{0.9}

      \resizebox{\linewidth}{!}{%
        \begin{tabular}{@{}llll@{}}
          \toprule
          \textbf{Primitive} &
          \textbf{Events} &
          \textbf{Coeffs} &
          \textbf{PTPX Row} \\
          \midrule

          PE tile
          & \texttt{pe\_tiles}
          & $\beta_{\mathtt{pe}}$
          & \texttt{PE\_tile} \\

          MEM tile
          & \texttt{mem\_tiles}
          & $\beta_{\mathtt{mem}}$
          & \texttt{MEM\_tile} \\

          REG
          & \texttt{registers}
          & $\beta_{\mathtt{reg}}$
          & \texttt{REG} \\

          IO tile
          & \texttt{IO\_tiles}
          & $\beta_{\mathtt{IO}}$
          & \texttt{IO\_tile} \\

          IC (SB)
          & \texttt{ic\_sb}
          & $\beta_{\mathtt{ic\_sb}}$
          & \texttt{SB} \\

          IC (RMUX)
          & \texttt{ic\_rmux}
          & $\beta_{\mathtt{ic\_rmux}}$
          & \texttt{RMUX} \\

          IC (PORT)
          & \texttt{ic\_port}
          & $\beta_{\mathtt{ic\_port}}$
          & \texttt{PORT} \\

          \bottomrule
        \end{tabular}%
      }
    \end{minipage}%
  }
  \hfill
  \subfloat[Power breakdown.\label{fig12-b}]{%
    \begin{minipage}[t]{0.42\columnwidth}
      \vspace{0pt}
      \centering

      \includegraphics[
        width=\linewidth,
        trim=0 0 0 0,
        clip
      ]{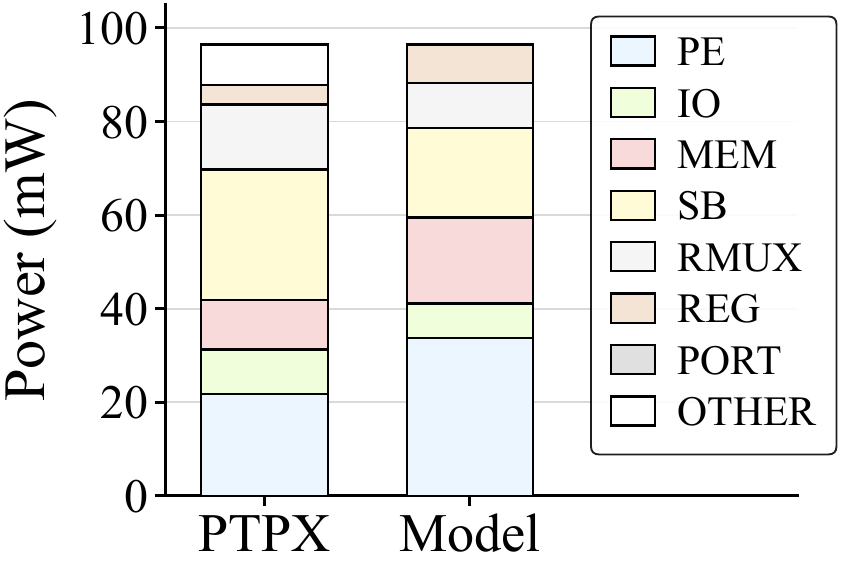}
    \end{minipage}%
  }

  \caption{Learned primitive coefficients and power breakdown.
  Panel (a) maps compiler-visible events to PTPX rows.
  Panel (b) compares Capstone and PTPX for VecAdd.}
  \label{fig12}
\end{figure}
\paragraph{Baselines}
Throughout the evaluation, we compare Capstone against several baselines. \emph{Cascade} is the unmodified compiler flow and selects post-PnR candidates using STA-reported frequency without compiler-level power prediction or cap enforcement. The \emph{analytical primitive} baseline uses manually specified or one-time-characterized costs for coarse compiler-visible primitive groups,
\[
P_{\mathrm{ana}}(x,f)
=
P_{\mathrm{leak}}
+
\rho(f)\sum_{r\in\mathcal{B}}\bar{e}_rN_r(x),
\]
where \(N_r(x)\) counts activity in primitive buckets such as PE, memory, register, switchbox, or routing-mux activity, and \(\bar{e}_r\) is its assigned cost. The \emph{scalar aggregate} baseline fits nonnegative event coefficients directly to total PTPX power using the same compiler-visible features, but does not use row-level physical supervision. The \emph{Capstone hierarchical} model is the deployed row-supervised compiler-resident predictor described in Section~\ref{sec-hier-energy-model}.

We additionally report a non-deployable \emph{same-abstraction oracle}. The oracle is restricted to the same compiler-visible event vocabulary as Capstone but is fit using privileged reference information that would not be available to a deployed compiler. Comparing Capstone with this oracle separates error introduced by model fitting from error inherent to the compiler-visible abstraction itself. Both predictors are evaluated against PTPX signoff power as the high-fidelity reference.

\subsection{Compiler-Resident Power Model}

\begin{figure}[t]
  \centering

  \subfloat[Run time breakdown.\label{fig13-a}]{%
    \begin{minipage}[t]{0.51\columnwidth}
      \vspace{0pt}
      \centering
      \scriptsize
      \setlength{\tabcolsep}{3pt}
      \renewcommand{\arraystretch}{0.9}

      \resizebox{\linewidth}{!}{%
        \begin{tabular}{@{}lcc@{}}
          \toprule
          \textbf{Component} &
          \textbf{Time / iter (s)} &
          \textbf{Share} \\
          \midrule

          STA (timing)
          & 2.68
          & 88.30\% \\

          Pipelining
          & 0.28
          & 9.30\% \\

          \textbf{Capstone predictor}
          & \textbf{0.01}
          & \textbf{$<\!1$\%} \\

          \midrule

          Post-PnR iter total
          & 3.04
          & 100\% \\

          \midrule

          Pipeline loop
          & 346.67
          & -- \\

          Signoff power
          & $1.07 \times 10^{5}$
          & -- \\

          \bottomrule
        \end{tabular}%
      }
    \end{minipage}%
  }
  \hfill
  \subfloat[Compile-time impact.\label{fig13-b}]{%
    \begin{minipage}[t]{0.47\columnwidth}
      \vspace{0pt}
      \centering

      \includegraphics[
        width=\linewidth
      ]{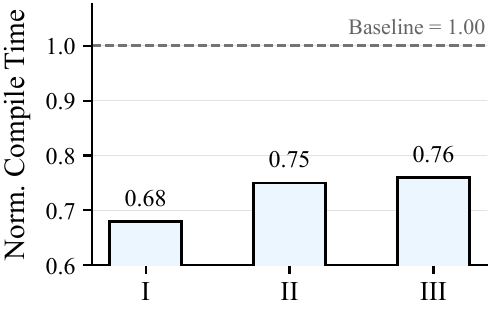}
    \end{minipage}%
  }

  \caption{Capstone run time impact for TTV.
  Panel (a) reports per-iteration run time.
  Panel (b) shows total compile time normalized to uncapped Cascade. Values below 1 arise because cap-aware stopping terminates post-PnR pipelining earlier despite the small per-iteration prediction overhead.}
  \label{fig13}
\end{figure}

We evaluate the compiler-resident power model along three dimensions needed for inner-loop power-aware optimization: prediction accuracy against signoff reference power, physical interpretability of the learned event-to-hardware decomposition, and overhead inside Cascade's post-PnR pipelining loop.

\paragraph{Prediction accuracy}
Figure~\ref{fig10} compares signoff PTPX power, Capstone prediction, and the same-abstraction oracle for each benchmark, with each kernel annotated by its Capstone absolute percentage error. Capstone achieves 16.8\% MAPE across the reported benchmark set, compared with 14.8\% for the oracle. Despite using privileged fitting information, the oracle improves MAPE only modestly. 

The per-kernel errors are not uniform, and several kernels exceed 30\% APE. This variation reflects information lost when gate-level switching, routing activity, local parasitics, and workload-dependent behavior are compressed into compiler-visible event activity. The relatively small gap between Capstone and the oracle suggests that a substantial portion of the remaining error comes from this abstraction boundary rather than only from the regression procedure. Therefore, we do not interpret the compiler-resident model as a replacement for signoff analysis. Our insight is that it provides a sufficiently fast physical proxy for compiler search when paired with the uncertainty margins in Section~\ref{sec-controllers}.

Table~\ref{tbl-model-baselines} provides a complementary comparison. The scalar aggregate model has substantially higher error than the row-supervised model, showing that total-power supervision alone does not sufficiently constrain event costs for this candidate set. The analytical primitive baseline has comparable aggregate accuracy, but requires manually specified or separately characterized primitive costs and does not learn Capstone's event-to-row physical attribution. In contrast, the hierarchical model learns both per-event costs and their mapping to physical power contributors directly from signoff supervision while exposing the event-level structure used by Capstone~III.

\paragraph{Physical attribution}
Beyond predicting total power, we evaluate whether the learned hierarchy recovers a meaningful relationship between compiler-visible events and gate-level power contributors. Figure~\ref{fig12}(a) reports the learned primitive-level coefficients and dominant PTPX rows associated with each compiler event group. The mapping aligns with architectural intuition. PE-related events concentrate on PE tile instances, memory events align with memory structures, and interconnect activity maps primarily to routing and register resources. Figure~\ref{fig12}(b) compares Capstone's primitive-level breakdown with PTPX aggregates for a representative kernel, showing that the sum-of-products predictor preserves the major physical contributions across PE, MEM, register, and interconnect structures. This decomposition also helps localize model error to particular primitive groups.

\paragraph{Compiler-loop overhead}
Finally, we measure the cost of evaluating the compiler-resident model inside Cascade's post-PnR optimization loop. Figure~\ref{fig13-a} reports the per-iteration run time breakdown for a representative kernel. STA dominates the iteration at 88.30\% of run time, while Cascade's pipelining pass accounts for 9.30\%. The Capstone predictor requires 0.01\,s, corresponding to $<$1\% per-iteration power-prediction overhead. The offline signoff power analysis requires approximately \(1.07\times10^{5}\)~s for the same design point, making the compiler-resident prediction roughly \(10^{7}\times\) faster. This gap explains why signoff analysis is useful as offline supervision but impractical as an inner-loop estimator.

\subsection{Compiler-Level Power-Aware Controllers}
We next evaluate the three controller modes from Section~\ref{sec-controllers}: Capstone~I with a static guardband, Capstone~II with a calibration-derived margin, and Capstone~III with bounded-error robust optimization. All three modes use the same compiler-resident power model and search the same post-PnR pipelining candidate space. They differ only in how they account for model uncertainty when deciding which bitstreams satisfy the power cap. We then verify the selected bitstreams using high-fidelity signoff power.


\begin{figure}[t]
  \centering
  \includegraphics[width=\linewidth]{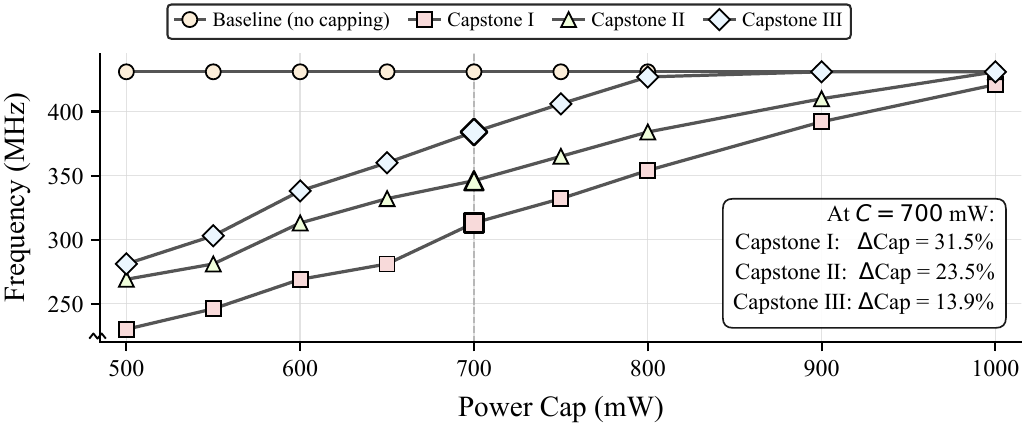}
\caption{Capstone controller evaluation for TTV. Maximum achievable frequency versus power cap for the baseline and Capstone I--III. $\Delta$Cap at 700 mW shows remaining headroom.}
  \label{fig12-combined}
\end{figure}
\paragraph{Controller configuration}
For Capstone~I, we use
\(\gamma_{\mathrm{spec}}=0.30\) and
\(\gamma_{\mathrm{anc}}=0.45\).
For Capstone~II, we target
\(\alpha_{\mathrm{anc}}=0.005\) and
\(\alpha_{\mathrm{spec}}=0.05\), with a held-out calibration set of
\(n_{\mathrm{global}}=24\) samples, including
\(n_{\mathrm{dense}}=9\) and \(n_{\mathrm{sparse}}=15\) samples in
candidate subgroups. Because \(\alpha_{\mathrm{anc}}=0.005\) would require about 199
calibration examples, our available \(n_{\mathrm{global}}=24\) samples are not enough to support a finite-sample 99.5\% conformal guarantee. In this evaluation, Capstone~II uses its conservative fallback margin instead. This still allows us to evaluate the controller's functionality under limited calibration data. Capstone~II uses held-out residuals to construct a data-driven margin, select bitstreams, and enforce the power cap, and we verify the selected bitstreams using signoff power. The results demonstrate the practical behavior and effectiveness of Capstone~II with the available calibration set, but they should not be interpreted as a new 99.5\% finite-sample coverage result. The formal conformal guarantee applies when enough calibration data are available and the required calibration assumptions hold.
We use
\[
\rho(f)=\max(1,f/f_{\mathrm{ref}})
\]
for residual frequency scaling. This choice prevents reducing the uncertainty margin above \(f_{\mathrm{ref}}\). For Capstone~III, event-level bounds combine model-fitting error, activity mismatch, PVT/DVFS margins, and deployment-shift margins, as summarized in Table~\ref{tbl-capstone-iii-sensitivity}.

\paragraph{Evaluation metrics}
We report signoff-verified success rate, returned bitstreams \(K\),
slack to cap, and normalized frequency. For selected signoff power
\(P\) under cap \(C\),
\[
\Delta\mathrm{Cap}
\triangleq
\frac{C-P}{C}\times100\%.
\]
A configuration is empirically feasible when \(\Delta\mathrm{Cap}\ge0\), while larger values indicate unused headroom. Normalized frequency is the selected frequency divided by the uncapped Cascade frequency for the corresponding kernel. The reported success metric is distinct from a controller's mode-specific guarantee. Success is determined after selection using signoff reference power, whereas the anchor guarantee depends on the guardband, calibration, or bounded-error assumptions in Section~\ref{sec-controllers}. In this evaluation, 100\% success means that all evaluated selections were below their caps at signoff.

\paragraph{Power cap sweep}
Figure~\ref{fig12-combined} illustrates controller behavior for TTV as the power cap varies. Frequency-only Cascade remains at the maximally pipelined configuration regardless of the cap, while Capstone selects different points from the post-PnR pipelining trajectory as the power budget changes. Capstone~I is most conservative, Capstone~II recovers more frequency under its margin, and Capstone~III reaches the highest frequency under the bounded-error assumption. As the cap is relaxed, all three modes approach the uncapped Cascade configuration. At 700\,mW, the figure reports the corresponding slack-to-cap values. These results show that Capstone can respond to the power budget by selecting a different bitstream before deployment, whereas throttling can only change the operating rate of an already selected implementation.

\begin{table}[t]
\centering
\caption{Aggregate controller metrics over kernel--cap pairs.}
\label{tbl-controller-metrics}
\small
\setlength{\tabcolsep}{3.2pt}
\renewcommand{\arraystretch}{1.02}

\begin{tabular*}{\columnwidth}{@{\extracolsep{\fill}}lcccc@{}}
\toprule
\textbf{Controller} &
\textbf{Success} &
\textbf{Norm. freq.} &
\textbf{Med. $\Delta$Cap} &
\textbf{$K$} \\
\midrule

Cascade &
0\% &
1.00 &
-- &
1 \\

Scalar Aggregate NNLS &
0\% &
1.00 &
-- &
1 \\

Signoff oracle &
100\% &
1.00 &
3.13\% &
1 \\

Capstone I &
100\% &
0.65 &
31.23\% &
$\leq 4$ \\

Capstone II &
100\% &
0.74 &
22.26\% &
$\leq 4$ \\

Capstone III, $K\leq4$ &
100\% &
0.81 &
13.85\% &
$\leq 4$ \\

\bottomrule
\end{tabular*}
\end{table}

\begin{figure}[t]
  \centering
  \includegraphics[width=\linewidth]{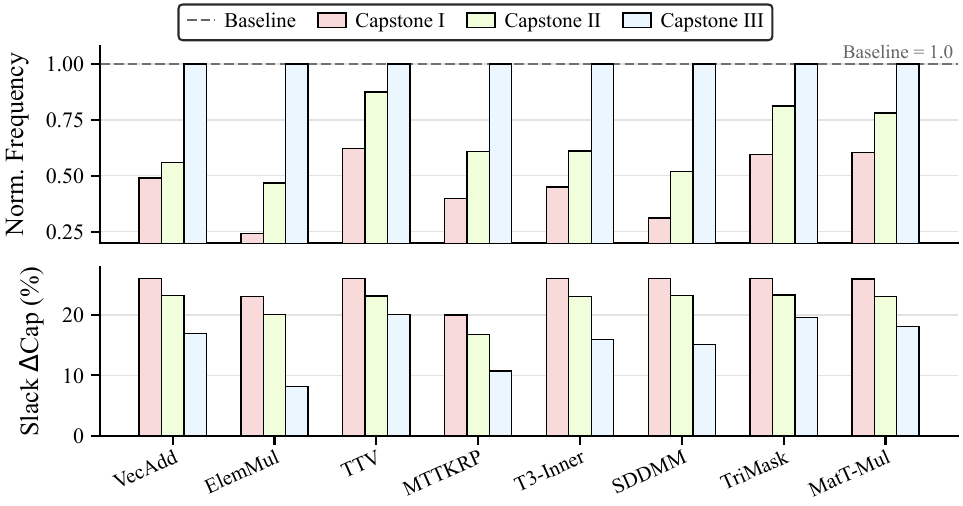}
\caption{Cross-kernel Capstone I--III controllers comparison. Top: normalized frequency relative to the uncapped baseline. Bottom: remaining slack to the power cap ($\Delta$Cap).}
  \label{fig-controller-cross-kernel}
\end{figure}
\begin{table}[t]
\centering
\caption{Capstone III sensitivity to bounded-error size and candidate-set pruning.}
\label{tbl-capstone-iii-sensitivity}
\footnotesize
\setlength{\tabcolsep}{2.0pt}
\renewcommand{\arraystretch}{1.02}

\resizebox{\columnwidth}{!}{%
\begin{tabular}{@{}lccccc@{}}
\toprule
\textbf{Setting} &
\textbf{Bound} &
\textbf{Success} &
\textbf{Norm. freq.} &
\textbf{Med. $\Delta$Cap} &
\textbf{$K$} \\
\midrule

$1\times$ fit &
$\epsilon_{\mathrm{fit}}$ &
100\% & 1.08 & 5.46\% & 90 \\

$2\times$ fit &
$2\epsilon_{\mathrm{fit}}$ &
100\% & 1.04 & 9.49\% & 90 \\

Fit + act. &
$\epsilon_{\mathrm{fit}}+\epsilon_{\mathrm{act}}$ &
100\% & 1.05 & 8.60\% & 90 \\

Fit + act. + PVT &
$+\epsilon_{\mathrm{PVT}}$ &
100\% & 1.02 & 11.82\% & 90 \\

Full bounds &
$+\epsilon_{\mathrm{OOD}}$ &
100\% & 1.00 & 13.85\% & 90 \\

\midrule

Unpruned &
Full bounds &
100\% & 1.00 & 13.85\% & 90 \\

Pruned, $K\leq8$ &
Top 8 retained &
100\% & 1.00 & 13.85\% & 8 \\

Pruned, $K\leq4$ &
Top 4 retained &
100\% & 1.00 & 13.85\% & 4 \\

\bottomrule
\end{tabular}%
}
\end{table}


\begin{table*}[t]
\centering
\caption{Capability-oriented comparison with prior CGRA compilers under target power caps.}
\label{tbl3}

\footnotesize
\setlength{\tabcolsep}{1.8pt}
\renewcommand{\arraystretch}{1.02}

\resizebox{\textwidth}{!}{%
\begin{tabular}{@{}l*{16}{c}@{}}
\toprule

\multicolumn{5}{c}{} &
\multicolumn{3}{c}{\textbf{Freq. (MHz)}} &
\multicolumn{3}{c}{\textbf{Power (mW)}} &
\multicolumn{3}{c}{\textbf{$\Delta$Cap (\%)}} &
\multicolumn{3}{c}{\textbf{Success}} \\

\cmidrule(lr){6-8}
\cmidrule(lr){9-11}
\cmidrule(lr){12-14}
\cmidrule(lr){15-17}

\textbf{Compiler} &
\textbf{Tech.} &
\textbf{Fabric} &
\textbf{Workload} &
\textbf{Cap (mW)} &

\textbf{Orig.} &
\textbf{2$\times$} &
\textbf{4$\times$} &

\textbf{Orig.} &
\textbf{2$\times$} &
\textbf{4$\times$} &

\textbf{Orig.} &
\textbf{2$\times$} &
\textbf{4$\times$} &

\textbf{Orig.} &
\textbf{2$\times$} &
\textbf{4$\times$} \\

\midrule

RipTide~\cite{gobieski-riptide-micro2022} &
22FFL &
6$\times$6 &
FFT &
0.30 &
50 &
25 &
12.5 &
0.24 &
0.12 &
0.06 &
20 &
60 &
80 &
Y &
Y &
Y \\

Snafu~\cite{gobieski-snafu-isca2021} &
22FFL &
6$\times$6 &
FFT &
0.40 &
50 &
25 &
12.5 &
0.54 &
0.27 &
0.135 &
$-35$ &
32.5 &
66.25 &
N &
Y &
Y \\

UE-CGRA~\cite{torng-uecgra-hpca2021} &
28nm &
8$\times$8 &
FFT &
5 &
750 &
325 &
162.5 &
14 &
7 &
3.5 &
$-180$ &
$-40$ &
30 &
N &
N &
Y \\

Plasticine~\cite{prabhakar-plasticine-isca2017} &
28nm &
-- &
Inner Prod. &
3000 &
280 &
140 &
70 &
18900 &
9450 &
4725 &
$-530$ &
$-215$ &
$-57.5$ &
N &
N &
N \\

Cascade~\cite{melchert-cascade-2024} &
12nm &
32$\times$16 &
Inner Prod. &
750 &
481 &
240.5 &
120.25 &
909.44 &
454.72 &
227.36 &
$-21.26$ &
39.37 &
69.69 &
N &
Y &
Y \\

Capstone I &
16nm &
32$\times$16 &
Inner Prod. &
750 &
294 &
147 &
73.5 &
510.24 &
255.12 &
127.56 &
31.97 &
65.98 &
82.99 &
Y &
Y &
Y \\

Capstone II &
16nm &
32$\times$16 &
Inner Prod. &
750 &
335 &
167.5 &
83.75 &
586.58 &
293.29 &
146.65 &
21.79 &
60.89 &
80.45 &
Y &
Y &
Y \\

Capstone III &
16nm &
32$\times$16 &
Inner Prod. &
750 &
366 &
183 &
91.5 &
645.57 &
322.78 &
161.39 &
13.92 &
56.96 &
78.48 &
Y &
Y &
Y \\

Cascade~\cite{melchert-cascade-2024} &
12nm &
32$\times$16 &
SDDMM &
1300 &
479 &
239.5 &
119.75 &
1488.56 &
744.28 &
372.14 &
$-14.5$ &
42.75 &
71.37 &
N &
Y &
Y \\

Capstone I &
16nm &
32$\times$16 &
SDDMM &
1300 &
310 &
155 &
77.5 &
893.98 &
446.99 &
223.5 &
31.23 &
65.62 &
82.81 &
Y &
Y &
Y \\

Capstone II &
16nm &
32$\times$16 &
SDDMM &
1300 &
346 &
173 &
86.5 &
1011.3 &
505.65 &
252.83 &
22.21 &
61.1 &
80.55 &
Y &
Y &
Y \\

Capstone III &
16nm &
32$\times$16 &
SDDMM &
1300 &
379 &
189.5 &
94.75 &
1121.44 &
560.72 &
280.36 &
13.74 &
56.87 &
78.43 &
Y &
Y &
Y \\

\bottomrule
\end{tabular}%
}
\end{table*}

\paragraph{Aggregate controller results}
Table~\ref{tbl-controller-metrics} summarizes feasibility, frequency, and headroom across the eight evaluated sparse kernels at their respective power caps. The uncapped Cascade baseline achieves 0\% success under these stressed caps because it performs no cap-aware selection. Under the evaluated margins, all three Capstone modes achieve 100\% cap compliance verified with signoff. Capstone~I, II, and III retain an average of 0.65\(\times\), 0.74\(\times\), and 0.81\(\times\) of the uncapped frequency, respectively, while leaving median cap slack of 31.23\%, 22.26\%, and 13.85\%. These results show the intended progression as Capstone~I leaves the most headroom, Capstone~II recovers additional performance with a tighter data-derived margin, and Capstone~III uses the available cap most aggressively under its explicit bounded-error assumption. For each controller, we preserve a conservative anchor and may expose speculative candidates when \(K>1\). Across the evaluated returned sets, speculation provides a 4.1\(\times\) frequency uplift relative to the anchor while reducing unused headroom by 24.6\% on average.

\paragraph{Cross-kernel robustness}
Figure~\ref{fig-controller-cross-kernel} shows the same pattern across the eight evaluated sparse vector, matrix, and tensor kernels. Capstone III achieves the highest normalized frequency, while Capstone II captures much of the improvement over the static guardband. Corresponding slack values show that these gains come from tighter use of the available budget while preserving empirical signoff feasibility.

Table~\ref{tbl-capstone-iii-sensitivity} isolates the effect of the Capstone~III uncertainty set. Starting from fit-only bounds, adding activity mismatch, PVT variation, and out-of-distribution margins increases the robust upper envelope, reducing achievable frequency and increasing cap slack. This is the expected tradeoff as larger uncertainty sets provide a more conservative condition for the robust anchor but leave more unused headroom. Restricting the returned set to \(K\le8\) or \(K\le4\) preserves the retained robust anchor and therefore its conditional guarantee.

\paragraph{Total compile-time impact}
Figure~\ref{fig13-b} reports total compile-time impact of the Capstone I--III controllers. Although Capstone adds a small amount of work to each post-PnR iteration, the compiler-resident predictor itself contributes less than 1\% of iteration time. Cap-aware stopping can also reduce total compilation time by terminating pipelining once additional structural optimization is no longer useful under the selected power bound. This shows how Capstone avoids an otherwise prohibitive signoff invocation inside the compiler loop without materially increasing per-iteration cost.

\subsection{Comparisons with State-of-the-Art CGRAs}
\label{sec-results-reliability}
Table~\ref{tbl3} presents a capability-oriented comparison with prior CGRA compilers under representative target caps. Because the designs differ in technology node, fabric size, workload, and reporting methodology, the table is not intended as an apples-to-apples architectural ranking. Instead, the comparison asks whether a flow can target a user-specified cap before deployment. Prior compilers do not perform compiler-level power capping, so we include optimistic throttling baselines that reduce each design's effective frequency and power by 2$\times$ and 4$\times$, modeling an even-duty-cycle run-time response~\cite{kim-arch-throttling-2006,lee-arch-throttling-2006,aragon-arch-throttling-2006}. These throttled points are optimistic because they assume power scales with the throttled execution rate and do not account for control overheads or residual static power.

The stressed caps in Table~\ref{tbl3} expose whether each flow can land on a feasible point under a fixed budget. Existing compilers may satisfy some caps incidentally, or only after coarse throttling of an already selected implementation. Capstone instead targets cap compliance during compilation as Capstone~I and II use conservative upper bounds, while Capstone~III searches over bounded-error candidates and returns at most \(K\) bitstreams with a guaranteed anchor. When the maximally pipelined design is already below the cap, Capstone~III can match the no-capping baseline. Under tighter caps, Capstone selects an earlier or structurally different post-PnR bitstream. Thus, the key distinction is that throttling changes the operating rate after deployment, whereas Capstone changes the bitstream before deployment, avoiding structural overhead from unnecessary over-pipelining.

Overall, Capstone provides a fast compiler-resident power estimate that
is accurate enough to guide post-PnR optimization when paired with
uncertainty-aware bounds. Across workloads, all three
controllers achieve 100\% empirical signoff-verified cap compliance,
while tighter uncertainty treatment recovers more frequency and reduces
unused power headroom. These results show that Capstone can enforce a
power budget by changing the selected bitstream before deployment.

\section{Conclusion}
\label{sec-conclusion}
We introduced Capstone, a compiler-level power-capping framework that combines an automated, data-driven methodology for learning compiler-visible event (energy) costs with uncertainty-aware post-PnR optimization. The compiler-resident power model approaches the accuracy of an oracle restricted to the same abstract event vocabulary while remaining fast enough for inner-loop use. Capstone's three controller modes then use this model to select hardware configurations under a power constraint while explicitly accounting for prediction uncertainty. In our evaluation, all three modes achieve full empirical signoff-verified cap compliance under the evaluated margins while recovering progressively more frequency with tighter uncertainty treatment. By acting at configuration time, Capstone can change the selected hardware configuration before deployment rather than relying only on run-time power management after the bitstream is fixed. We hope this work lays the foundation for compilers that treat energy and power as a first-class optimization objective alongside performance.

\section{Acknowledgements}

This work was supported in part by a gift from Intel Corporation, and the NSF Graduate Research Fellowship Program (GRFP) Award \#2236421. We thank Intel for providing access to the Intel 16 PDK.

\newpage
\bibliographystyle{IEEEtran}
\bibliography{main}

@string{JAN = {Jan}}

@string{FEB = {Feb}}

@string{MAR = {Mar}}

@string{APR = {Apr}}

@string{JUN = {Jun}}

@string{JUL = {Jul}}

@string{SEP = {Sep}}

@string{OCT = {Oct}}

@string{NOV = {Nov}}

@string{DEC = {Dec}}

@string{PLDI     = {ACM SIGPLAN Conf. on Programming Language Design
                     and Implementation (PLDI)}}

@string{TCAD     = {IEEE Trans. on Computer-Aided Design of Integrated
                     Circuits and Systems (TCAD)}}

@string{TECS     = {IEEE Trans. on Embedded Computing Systems (TECS)}}

@string{TACO     = {ACM Trans. on Architecture and Code
                     Optimization (TACO)}}

@string{COMPUTER = {IEEE Computer}}

@string{ISCA     = {Int'l Symp. on Computer Architecture (ISCA)}}

@string{MICRO    = {Int'l Symp. on Microarchitecture (MICRO)}}

@string{HPCA     = {Int'l Symp. on High-Performance Computer
                     Architecture (HPCA)}}

@string{ASPLOS   = {Int'l Conf. on Architectural Support for Programming
                     Languages and Operating Systems (ASPLOS)}}

@string{FPGA     = {Int'l Symp. on Field Programmable Gate Arrays (FPGA)}}

@string{TODAES   = {ACM Trans. on Design Automation
                     of Electronic Systems (TODAES)}}

@string{ISQED    = {Int'l Symp. on Quality Electronic Design (ISQED)}}

@string{DAC      = {Design Automation Conf. (DAC)}}

@string{DATE     = {Design, Automation, and Test in Europe (DATE)}}

@string{ICCD     = {Int'l Conf. on Computer Design (ICCD)}}

@string{ASPDAC   = {Asia and South Pacific Design Automation
                     Conference (ASPDAC)}}

@string{ICCAD    = {Int'l Conf. on Computer-Aided Design (ICCAD)}}

@string{TVLSI    = {IEEE Trans. on Very Large-Scale Integration Systems (TVLSI)}}

@article{torng-uecgra-hpca2021,
  title     = {Ultra-Elastic CGRAs for Irregular Loop Specialization},
  author    = {Torng, Christopher and Peitian Pan and Yanghui Ou and Cheng Tan and Christopher Batten},
  journal   = HPCA,
  month     = FEB,
  year      = {2021},
}

@article{li-mcpat-taco2013,
  title     = {The McPAT Framework for Multicore and Manycore
                Architectures: Simultaneously Modeling Power, Area, and
                Timing},
  author    = {Sheng Li and Jung Ho Ahn and Richard D. Strong and Jay B.
                Brockman and Dean M. Tullsen and Norman P. Jouppi},
  journal   = TACO,
  month     = APR,
  year      = {2013},
  volume    = {10},
  number    = {1},
  pages     = {5:1--5:29},
}

@article{prabhakar-plasticine-isca2017,
  title     = {Plasticine: A Reconfigurable Architecture For Parallel Patterns},
  author    = {Raghu Prabhakar and Yaqi Zhang and David Koeplinger and
                Matthew Feldman and Tian Zhao and Stefan Hadjis and
                Ardavan Pedram and Christos Kozyrakis and Kunle Olukotun},
  journal   = ISCA,
  month     = JUN,
  year      = {2017},
}

@article{huang-elastic-cgras-fpga2013,
  title     = {Elastic CGRAs},
  author    = {Huang, Yuanjie and Ienne, Paolo and Temam, Olivier and Chen, Yunji and Wu, Chengyong},
  journal   = FPGA,
  month     = FEB,
  year      = {2013},
}

@article{vasilyev-vision-micro2016,
  title     = {Evaluating programmable architectures for imaging and
                vision applications},
  author    = {Artem Vasilyev and Nikhil Bhagdikar and Ardavan Pedram and
                Stephen Richardson and Shahar Kvatinsky and Mark
                  Horowitz},
  journal   = MICRO,
  month     = OCT,
  year      = {2016},
}

@article{fan-cgra-objectinf-tvlsi2018,
  title     = {Stream Processing Dual-Track CGRA for Object Inference},
  author    = {Xitian Fan and Di Wu and Wei Cao and Wayne Luk and Lingli Wang},
  journal   = TVLSI,
  month     = FEB,
  year      = {2018},
}

@article{kim-samsung-cgra-fpt2012,
  title     = {ULP-SRP: Ultra low power Samsung Reconfigurable Processor
                for biomedical applications},
  author    = {C. {Kim} and M. {Chung} and Y. {Cho} and M. {Konijnenburg}
                and S. {Ryu} and J. {Kim}},
  journal   = {International Conference on Field-Programmable Technology
                (FPT)},
  month     = DEC,
  year      = {2012},
}

@article{jafri-cgra-isqed2013,
  title     = {Energy-aware coarse-grained reconfigurable architectures
                using dynamically reconfigurable isolation cells},
  author    = {S. M. A. H. {Jafri} and O. {Bag} and A. {Hemani} and N.
                {Farahini} and K. {Paul} and J. {Plosila} and H.
                {Tenhunen}},
  journal   = ISQED,
  month     = MAR,
  year      = {2013},
}

@article{karunaratne-hycube-dac2017,
  title   = {HyCUBE: A CGRA with Reconfigurable Single-cycle Multi-hop
              Interconnect},
  author  = {Manupa Karunaratne and Aditi Kulkarni Mohite and Tulika Mitra
              and Li-Shiuan Peh},
  journal = DAC,
  month   = JUN,
  year    = {2017},
}

@article{koul-aha-tecs2023,
  title     = {AHA: An Agile Approach to the Design of Coarse-Grained Reconfigurable Accelerators and Compilers},
  author    = {Kalhan Koul and Jackson Melchert and Kavya Sreedhar and Leonard Truong and Gedeon Nyengele and Keyi Zhang and Qiaoyi Liu and Jeff Setter and Po-Han Chen and Yuchen Mei and Maxwell Strange and Ross Daly and Caleb Donovick and Alex Carsello and Taeyoung Kong and Kathleen Feng and Dillon Huff and Ankita Nayak and Rajsekhar Setaluri and James Thomas and Nikhil Bhagdikar and David Durst and Zachary Myers and Nestan Tsiskaridze and Stephen Richardson and Rick Bahr and Kayvon Fatahalian and Pat Hanrahan and Clark Barrett and Mark Horowitz and Christopher Torng and Fredrik Kjolstad and Priyanka Raina},
  journal   = TECS,
  month     = JAN,
  year      = {2023},
}

@article{gobieski-riptide-micro2022,
  title     = {RipTide: A Programmable, Energy-Minimal Dataflow Compiler and Architecture},
  author    = {Gobieski, Graham and Ghosh, Souradip and Heule, Marijn and Mowry, Todd and Nowatzki, Tony and Beckmann, Nathan and Lucia, Brandon},
  journal   = MICRO,
  month     = OCT,
  year      = {2022},
}

@article{gobieski-snafu-isca2021,
  title     = {Snafu: An Ultra-Low-Power, Energy-Minimal CGRA-Generation Framework and Architecture},
  author    = {Gobieski, Graham and Atli, Ahmet Oguz and Mai, Kenneth and Lucia, Brandon and Beckmann, Nathan},
  journal   = ISCA,
  month     = JUN,
  year      = {2021},
}

@article{hsu-sparse-sam-asplos2023,
  title     ={The sparse abstract machine}, 
  author    ={Hsu, Olivia and Maxwell Strange and Ritvik Sharma and Jaeyeon Won and Kunle Olukotun and Joel S. Emer and Mark A. Horowitz and Fredrik Kjølstad},
  journal   = ASPLOS,
  month     = MAR,
  year      = {2023},
}

@article{melchert-cascade-2024,
  title = {Cascade: An Application Pipelining Toolkit for Coarse-Grained Reconfigurable Arrays}, 
  author = {Melchert, Jackson and Mei, Yuchen and Koul, Kalhan and Liu, Qiaoyi and Horowitz, Mark and Raina, Priyanka},
  journal   = TCAD,
  month     = OCT,
  year      = {2024},
}

@article{zoni-runtime-edge-power-monitors-2023,
  title = {A Survey on Run-time Power Monitors at the Edge},
  author = {Zoni, Davide and Galimberti, Andrea and Fornaciari, William},
  journal   = {ACM Comput. Surv.},
  month     = JUL,
  year      = {2023},
}

@article{zoni-energy-constrained-controller-2020,
  title = {All-Digital Energy-Constrained Controller for General-Purpose Accelerators and CPUs}, 
  author = {Zoni, Davide and Cremona, Luca and Fornaciari, William},
  journal   = {IEEE Embedded Systems Letters}, 
  month     = MAR,
  year      = {2020},
}

@article{xu-os-2015,
  title = {Automated OS-level Device Runtime Power Management},
  author = {Xu, Chao and Lin, Felix Xiaozhu and Wang, Yuyang and Zhong, Lin},
  journal   = ASPLOS, 
  month     = MAR,
  year      = {2015},
}

@article{chen-energy-2015,
  title = {Execution time prediction for energy-efficient hardware accelerators},
  author = {Chen, Tao and Rucker, Alexander and Suh, G. Edward},
  journal   = MICRO, 
  month     = OCT,
  year      = {2015},
}

@article{tang-neurometer-2021,
  title = {NeuroMeter: An Integrated Power, Area, and Timing Modeling Framework for Machine Learning Accelerators Industry Track Paper},
  author = {Tang, Tianqi and Li, Sheng and Nai, Lifeng and Jouppi, Norm and Xie, Yuan},
  journal   = HPCA, 
  month     = FEB,
  year      = {2021},
}

@article{zoni-powerprobe-2018,
  title = {PowerProbe: Run-time power modeling through automatic RTL instrumentation}, 
  author = {Zoni, Davide and Cremona, Luca and Fornaciari, William},
  journal   = DATE, 
  month     = MAR,
  year      = {2018},
}

@article{nabavinejad-batching-dvfs-2022,
  title = {Coordinated Batching and DVFS for DNN Inference on GPU Accelerators}, 
  author = {Nabavinejad, Seyed Morteza and Reda, Sherief and Ebrahimi, Masoumeh},
  journal   = {IEEE Transactions on Parallel and Distributed Systems}, 
  month     = JAN,
  year      = {2022},
}

@article{liu-runtime-2013,
  title = {Achieving energy efficiency through runtime partial reconfiguration on reconfigurable systems}, 
  author = {Liu, Shaoshan and Pittman, Richard Neil and Forin, Alessandro and Gaudiot, Jean-Luc},
  journal   = {ACM Trans. Embed. Comput. Syst.},
  month     = MAR,
  year      = {2013},
}

@article{tann-runtime-config-2016,
  title = {Runtime configurable deep neural networks for energy-accuracy trade-off},
  author = {Tann, Hokchhay and Hashemi, Soheil and Bahar, R. Iris and Reda, Sherief},
  journal   = {IEEE/ACM/IFIP Int'l Conference on Hardware/Software Codesign and System Synthesis (CODES)},
  month     = OCT,
  year      = {2016},
}

@article{nabavinejad-batchsizer-2021,
  title = {BatchSizer: Power-Performance Trade-off for DNN Inference},
  author = {Nabavinejad, Seyed Morteza and Reda, Sherief and Ebrahimi, Masoumeh},
  journal   = {Proceedings of the 26th Asia and South Pacific Design Automation Conference (ASPDAC)}, 
  month     = JAN,
  year      = {2021},
}

@article{kim-simmani-2019,
  title = {Simmani: Runtime Power Modeling for Arbitrary RTL with Automatic Signal Selection},
  author = {Kim, Donggyu and Zhao, Jerry and Bachrach, Jonathan and Asanovi\'{c}, Krste},
  journal   = MICRO, 
  month     = OCT,
  year      = {2019},
}

@article{xu-dynamic-2017,
  title = {Approximate Reconfigurable Hardware Accelerator: Adapting the Micro-Architecture to Dynamic Workloads}, 
  author = {Xu, Siyuan and Schafer, Benjamin Carrion},
  journal   = ICCD, 
  month     = NOV,
  year      = {2017},
}

@article{nasser-power-2021,
  title = {RTL to Transistor Level Power Modeling and Estimation Techniques for FPGA and ASIC: A Survey},
  author = {Nasser, Yehya and Lorandel, Jordane and Prévotet, Jean-Christophe and Hélard, Maryline},
  journal   = {IEEE Transactions on Computer-Aided Design of Integrated Circuits and Systems}, 
  month     = JUN,
  year      = {2021},
}

@article{lin-hl-pow-2020,
  title = {HL-Pow: A Learning-Based Power Modeling Framework for High-Level Synthesis},
  author = {Lin, Zhe and Zhao, Jieru and Sinha, Sharad and Zhang, Wei},
  journal   = {IEEE Asia and South Pacific Design Automation Conference (ASP-DAC)}, 
  month     = JAN,
  year      = {2020},
}

@article{kumar-cpu-power-2023,
  title = {Machine Learning-Based Microarchitecture- Level Power Modeling of CPUs},
  author = {Kumar, Ajay Krishna Ananda and Al-Salamin, Sami and Amrouch, Hussam and Gerstlauer, Andreas},
  journal   = {IEEE Transactions on Computers}, 
  month     = APR,
  year      = {2023},
}

@article{lee-power-model-2018,
  title = {Learning-Based, Fine-Grain Power Modeling of System-Level Hardware IPs},
  author = {Lee, Dongwook and Gerstlauer, Andreas},
  journal   = {ACM Transactions on Design Automation of Electronic Systems (TODAES)}, 
  month     = FEB,
  year      = {2018},
}

@article{zhou-primal-2019,
  title = {PRIMAL: Power Inference using Machine Learning},
  author = {Zhou, Yuan and Ren, Haoxing and Zhang, Yanqing and Keller, Ben and Khailany, Brucek and Zhang, Zhiru},
  journal   = DAC, 
  month     = JUN,
  year      = {2019},
}

@article{coburn-power-2005,
  title = {Power emulation: a new paradigm for power estimation},
  author = {Coburn, Joel and Ravi, Srivaths and Raghunathan, Anand},
  journal   = DAC, 
  month     = JUN,
  year      = {2005},
}

@article{zhang-grannite-2020,
  title = {GRANNITE: Graph Neural Network Inference for Transferable Power Estimation},
  author = {Zhang, Yanqing and Ren, Haoxing and Khailany, Brucek},
  journal   = DAC, 
  month     = JUL,
  year      = {2020},
}

@article{jianlei-power-2015,
  title = {Early stage real-time SoC power estimation using RTL instrumentation},
  author = {Jianlei Yang and Liwei Ma and Kang Zhao and Yici Cai and Tin-Fook Ngai},
  journal   = {Asia and South Pacific Design Automation Conference (ASPDAC)}, 
  month     = JAN,
  year      = {2015},
}

@article{kim-strober-2016,
  title = {Strober: Fast and Accurate Sample-Based Energy Simulation for Arbitrary RTL},
  author = {Donggyu Kim and Adam Izraelevitz and Christopher Celio and Hokeun Kim and Brian Zimmer and Yunsup Lee and Jonathan Bachrach and Krste Asanovi{\'c}},
  journal   = {International Symposium on Computer Architecture (ISCA)}, 
  month     = jun,
  year      = {2016},
}

@article{rahimi-guardband-2013,
  title = {Hierarchically focused guardbanding: an adaptive approach to mitigate PVT variations and aging},
  author = {Rahimi, Abbas and Benini, Luca and Gupta, Rajesh K.},
  journal   = DATE, 
  month     = MAR,
  year      = {2013},
}

@article{rahimi-guardband-2014,
  title = {Application-Adaptive Guardbanding to Mitigate Static and Dynamic Variability},
  author = {Rahimi, Abbas and Benini, Luca and Gupta, Rajesh K.},
  journal   = {IEEE Transactions on Computers}, 
  month     = SEP,
  year      = {2014},
}

@article{srivastava-robust-opt-2007,
  title = {Low-Power-Design Space Exploration Considering Process Variation Using Robust Optimization},
  author = {Srivastava, Ashish and Kachru, Tejasvi and Sylvester, Dennis},
  journal   = {IEEE Transactions on Computer-Aided Design of Integrated Circuits and Systems}, 
  month     = JAN,
  year      = {2007},
}

@article{zhang-chance-constraints-2011,
  title = {Chance Constrained Programming for Optimal Power Flow Under Uncertainty},
  author = {Zhang, Hui and Li, Pu},
  journal   = {IEEE Transactions on Power Systems}, 
  month     = NOV,
  year      = {2011},
}

@article{duan-chance-constraints-2018,
  title = {Distributionally Robust Chance-Constrained Approximate AC-OPF With Wasserstein Metric},
  author = {Duan, Chao and Fang, Wanliang and Jiang, Lin and Yao, Li and Liu, Jun},
  journal   = {IEEE Transactions on Power Systems}, 
  month     = SEP,
  year      = {2018},
}

@article{cocheteux-conformal-pred-calib-2025,
  title   = {Uncertainty-Aware Online Extrinsic Calibration: A Conformal Prediction Approach},
  author  = {Cocheteux, Mathieu and Moreau, Julien and Davoine, Franck},
  journal = {IEEE/CVF Winter Conference on Applications of Computer Vision (WACV)},
  month   = FEB,
  year    = {2025}
}

@article{cohen-conformal-pred-calib-2025,
  title   = {Calibrating AI Models for Wireless Communications via Conformal Prediction},
  author  = {Cohen, Kfir M. and Park, Sangwoo and Simeone, Osvaldo and Shamai Shitz, Shlomo},
  journal = {IEEE Transactions on Machine Learning in Communications and Networking},
  month   = SEP,
  year    = {2023}
}

@article{yeh-conformal-pred-calib-2024,
  title = {End-to-End Conformal Calibration for Optimization Under Uncertainty},
  author = {Christopher Yeh and Nicolas Christianson and Alan Wu and Adam Wierman and Yisong Yue},
  journal   = {Transactions on Machine Learning Research}, 
  month     = dec,
  year      = {2025},
}

@article{gopakumar-conformal-pred-calib-2024,
  title = {Uncertainty Quantification of Surrogate Models Using Conformal Prediction},
  author = {Vignesh Gopakumar and Ander Gray and Joel Oskarsson and Lorenzo Zanisi and Stanislas Pamela and Daniel Giles and Matt J. Kusner and Marc Peter Deisenroth},
  journal   = {Machine Learning: Science and Technology}, 
  month     = feb,
  year      = {2026},
}

@article{lu-conformal-pred-calib-2023,
  title   = {Federated Conformal Predictors for Distributed Uncertainty Quantification},
  author  = {Lu, Charles and Yu, Yaodong and Karimireddy, Sai Praneeth and Jordan, Michael and Raskar, Ramesh},
  journal = {Proceedings of the 40th International Conference on Machine Learning (ICML), PMLR 202:22942--22964},
  month   = JUL,
  year    = {2023}
}

@article{aragon-arch-throttling-2006,
  title   = {Power-aware control speculation through selective throttling},
  author  = {Aragon, J.L. and Gonzalez, J. and Gonzalez, A.},
  journal = HPCA,
  month   = FEB,
  year    = {2003}
}

@article{kim-arch-throttling-2006,
  title   = {Performance boosting under reliability and power constraints},
  author  = {Kim, Youngtaek and John, Lizy Kurian and Paul, Indrani and Manne, Srilatha and Schulte, Michael},
  journal = ICCAD,
  month   = NOV,
  year    = {2013}
}

@article{lee-arch-throttling-2006,
  title   = {Throttling-Based Resource Management in High Performance Multithreaded Architectures},
  author  = {Lee, Seong-Won and Gaudiot, J.-L.},
  journal = {IEEE Transactions on Computers},
  month   = SEP,
  year    = {2006}
}

@article{reddi-arch-throttling-2009,
  title   = {Voltage Emergency Prediction: Using Signatures to Reduce Operating Margins},
  author  = {Reddi, Vijay Janapa and Gupta, Meeta S. and Holloway, Glenn and Wei, Gu-Yeon and Smith, Michael D. and Brooks, David},
  journal = HPCA,
  month   = FEB,
  year    = {2009}
}

@misc{StanfordAHA_garnet,
  author       = {{Stanford AHA! Agile Hardware Center}},
  title        = {garnet: Next generation CGRA generator},
  year         = {2019},
  howpublished = {GitHub repository},
  url          = {https://github.com/StanfordAHA/garnet}
}

@article{daly-coreir-2018,
  title   = {Invoking and Linking Generators from Multiple Hardware
Languages using CoreIR},
  author  = {Daly, Ross and Truong, Leonard and Hanrahan, Pat},
  journal = {In Workshop on Open-Source EDA Technology (WOSET)},
  month   = SEP,
  year    = {2018}
}

@article{ragan-halide-2013,
  title = {Halide: A Language and Compiler for Optimizing Parallelism, Locality, and Recomputation in Image Processing Pipelines},
  author = {Jonathan Ragan-Kelley and Connelly Barnes and Andrew Adams and Sylvain Paris and Fr{\'e}do Durand and Saman Amarasinghe},
  journal   = {Programming Language Design and Implementation (PLDI)}, 
  month     = jun,
  year      = {2013},
}

@article{huff-clockwork-2021,
  title   = {Clockwork: Resource-Efficient Static Scheduling for Multi-Rate Image Processing Applications on FPGAs},
  author  = {Huff, Dillon and Dai, Steve and Hanrahan, Pat},
  journal = {Int'l Symposium on Field-Programmable Custom Computing Machines},
  month   = Jun,
  year    = {2021}
}

@inproceedings{he-hybrid-gpufpga-ics2021,
author    = {Xin He and Jiawen Liu and Zhen Xie and Hao Chen and Guoyang Chen and Weifeng Zhang and Dong Li},
title     = {Enabling Energy-Efficient {DNN} Training on Hybrid {GPU-FPGA} Accelerators},
booktitle = {Proceedings of the 35th {ACM} International Conference on Supercomputing},
pages     = {227--241},
year      = {2021},
doi       = {10.1145/3447818.3460371}
}

@article{bertsimas-sim-robustness-2004,
  title     = {The Price of Robustness},
  author    = {Dimitris Bertsimas and Melvyn Sim},
  journal   = {Operations Research},
  year      = {2004},
  volume    = {52},
  number    = {1},
  pages     = {35--53},
  doi       = {10.1287/opre.1030.0065},
}

@article{gu-multivdd-cgra-aspdac2017,
  title     = {Energy-Aware Loops Mapping on Multi-{VDD} {CGRAs} without Performance Degradation},
  author    = {Jiangyuan Gu and Shouyi Yin and Leibo Liu and Shaojun Wei},
  journal   = ASPDAC,
  month     = JAN,
  year      = {2017},
}

@article{kojima-body-bias-cgra-ieice2018,
  title     = {Optimization of Body Biasing for Variable Pipelined Coarse-Grained Reconfigurable Architectures},
  author    = {Takuya Kojima and Naoki Ando and Hayate Okuhara and Ng. Anh Vu Doan and Hideharu Amano},
  journal   = {IEICE Transactions on Information and Systems},
  month     = JUN,
  year      = {2018},
  volume    = {E101-D},
  number    = {6},
  pages     = {1532--1540},
  doi       = {10.1587/transinf.2017EDP7308},
}

@article{nalesh-energy-aware-cgra-integration2017,
  title     = {Energy Aware Synthesis of Application Kernels through Composition of Data-Paths on a {CGRA}},
  author    = {Nalesh S. and Kavitha T. Madhu and Saptarsi Das and S. K. Nandy and Ranjani Narayan},
  journal   = {Integration},
  year      = {2017},
  volume    = {58},
  pages     = {320--328},
  doi       = {10.1016/j.vlsi.2017.02.009},
}

@article{nemirovski-shapiro-chance-2006,
  title     = {Convex Approximations of Chance Constrained Programs},
  author    = {Arkadi Nemirovski and Alexander Shapiro},
  journal   = {SIAM Journal on Optimization},
  year      = {2006},
  volume    = {17},
  number    = {4},
  pages     = {969--996},
  doi       = {10.1137/050622328},
}

@article{peng-battery-aware-cgra-aspdac2015,
  title     = {Battery-Aware Mapping Optimization of Loop Nests for {CGRAs}},
  author    = {Yu Peng and Shouyi Yin and Leibo Liu and Shaojun Wei},
  journal   = ASPDAC,
  month     = JAN,
  year      = {2015},
}

@article{shafer-vovk-conformal-2008,
  title     = {A Tutorial on Conformal Prediction},
  author    = {Glenn Shafer and Vladimir Vovk},
  journal   = {Journal of Machine Learning Research},
  year      = {2008},
  volume    = {9},
  number    = {12},
  pages     = {371--421},
}

\clearpage
\appendix
\section{Artifact Appendix}

\subsection{Abstract}
This artifact provides the source code for Capstone's hierarchical power modeling methodology and compiler-level power-aware controllers, along with the scripts used to reproduce the released evaluation results. It includes the deployed compiler-side power model, Capstone I--III controllers, Cascade integration, traces from the eight evaluated kernels, and plotting scripts. The artifact reproduces Figures 5, 9, and 11--13 and Tables VII--IX, and generates functional versions of Figures 5, 9, and 10 using deterministic synthetic power reports. The original Synopsys \ptpx{} (PTPX) hierarchy reports and row-level power values are protected under NDA and cannot be distributed.

We provide both a Quick Start workflow and a comprehensive end-to-end evaluation flow that integrates the Cascade CGRA compiler. The Quick Start reproduces the published figures and tables in under two minutes after dependencies are installed. This workflow only requires a local Python environment and serves as a rapid validation of the installation and output pipeline. The complete end-to-end flow begins in the AHA Tutorial Docker container, where it installs the released pipeline, STA, RTL export, coefficient, and bound files before running each application through the post-place-and-route pipelining loop to evaluate the Baseline and Capstone I--III controllers. This complete workflow takes approximately 90 minutes and also generates functional versions of the hierarchical power model results. The Quick Start can be evaluated on a machine with 8 GB of memory, and 5 GB of free disk space, while the comprehensive flow requires an x86-64 Linux host with eight CPU cores, 32 GB of RAM, and 30 GB of free disk space.

\subsection{Artifact check-list (meta-information)}

{\small
\begin{itemize}
  \setlength{\itemsep}{0pt}
  \item {\bf Algorithms:} hierarchical nonnegative fitting with architecture-informed event-to-row constraints, activity-proxy prediction, and Capstone I--III controllers.
  \item {\bf Programs:} Python scripts; modified AHA pipeline, STA, RTL export, and plotting.
  \item {\bf Compilation:} Python; AHA Docker; SAM graphs; Cascade post-PnR backend.
  \item \textbf{Transformations:} interconnect-register insertion; selection of up to \(K=4\) configurations.
  \item \textbf{Binary:} external
  \path{stanfordaha/garnet:micro-demos} Docker image.
  \item {\bf Models:} dynamic/leakage coefficients, activity proxy, and Capstone III bounds.
  \item {\bf Data sets:} eight kernel traces, tensor3-ttv cap sweep, and synthetic reports.
  \item {\bf Run-time environment:} Python 3 on Linux/macOS; Docker for AHA regeneration.
  \item {\bf Hardware:} Quick Start: 2 CPU cores, 8\,GB RAM; full flow recommended: x86-64 Linux, 8 cores, 32\,GB RAM.
  \item {\bf Run-time state: } fixed Quick Start inputs; environment-configured Docker runs; reused \path{SIM_DIR}.
  \item {\bf Execution: } two Quick Start commands; nine AHA runs for full regeneration.
  \item {\bf Metrics:} percent error, MAPE, \(R^2\), frequency, cap slack, success, and time.
  \item {\bf Output:} PDF/PNG figures and CSV/JSON results.
  \item {\bf Experiments:} Figures 5, 9, and 11--13 and Tables VII--IX; synthetic Figures 5, 9, and 10.
  \item {\bf Disk Space: } 5 GB Quick Start; 30 GB full Docker flow.
  \item {\bf Time:} 5--10 minute setup; \(<2\) minute Quick Start; approximately 90 minute full flow.
  \item {\raggedright
  {\bf Publicly available:}
  \url{https://github.com/usc-acorn/yarzada-capstone-micro2026}.\par}
  \item {\bf Workflow automation:} two commands generate and validate all public outputs in Quick Start.
  \item {\bf Archived version:} \url{https://doi.org/10.5281/zenodo.21880340}.
\end{itemize}
}
\subsection{Description}

\subsubsection{How to access}

Clone the public repository at
\url{https://github.com/usc-acorn/yarzada-capstone-micro2026}. The repository README
provides the complete command reference and expected outputs.

\subsubsection{Hardware dependencies}

The Quick Start requires 8\,GB RAM and
5\,GB of free disk space. For the comprehensive Docker workflow, we recommend
an x86-64 Linux host with eight CPU cores, 32\,GB RAM, and 30\,GB of free disk
space.

\subsubsection{Software dependencies}

The Quick Start requires Python 3.9 or newer and the packages listed in
\texttt{requirements.txt}. Controller-trace regeneration additionally requires
Docker and the AHA image \path{stanfordaha/garnet:micro-demos}.

\subsubsection{Data sets}

Each \path{data/capstone_<kernel>/} directory contains the candidate trace,
controller and bitstream summaries, timing measurements, and selection
manifest for one evaluated kernel. The
\path{data/capstone_tensor3_ttv_high_cap/} directory contains the complete
trajectory used to construct the Figure 12 sweep, and
\path{data/figure12_sweep.csv} contains the derived sweep results.
\path{data/power_model_synthetic/} contains deterministic synthetic hierarchy
reports, compiler-event counts, and train/held-out split labels.

\subsubsection{Models}

\path{coeffs/} contains the learned coefficients, activity-proxy fit, and
Capstone III bounds. As in the paper, the deployed model is
\[
  \mu(x)=\sum_{e=1}^{E}\beta_e X_e(x)+P_{\mathrm{leak}},
  \qquad
  \beta_e=\alpha_e\sum_{r=1}^{R}W_{re}.
\]
The public synthetic fit uses architecture-informed event-to-row constraints to regularize limited, correlated training data while learning the nonnegative allocation magnitudes and event costs. \texttt{pipeline.py} converts compiler event counts to activity using frequency and initiation interval scaling and computes \(P_{\mathrm{leak}}\) from the released leakage coefficients.

\subsection{Installation}

From the repository root:
{\footnotesize
\begin{verbatim}
python3 -m venv .venv
source .venv/bin/activate
python3 -m pip install -r requirements.txt
\end{verbatim}
}

\subsection{Experiment workflow}

\subsubsection{Quick Start}

Using the bundled data, run:
{\footnotesize
\begin{verbatim}
python3 src/capstone_power_model.py evaluate \
  --manifest data/power_model_synthetic/manifest.json \
  --output-dir figures/generated_power_model \
  --overwrite --validate-reference

python3 plot/generate_capstone_figures.py \
  --data-dir data \
  --output-dir figures \
  --validate-reference
\end{verbatim}
}

The first command generates synthetic functional demonstrations corresponding to Figures 5, 9, and 10 under \path{figures/generated_power_model/}. The second generates the released paper/reference versions of Figures 5 and 9, along with Figures 11--13 and Tables VII--IX, directly under \path{figures/}. Validation against the bundled references reports whether the workflow succeeded.

\subsubsection{End-to-end controller flow}

Follow README Sections 1--2 to install the released AHA files, select a kernel and its documented cap, and run:
{\footnotesize
\begin{verbatim}
docker exec -it capstone bash
cd /aha
./cascade_demo.sh max
\end{verbatim}
}
The pipeline evaluates the Baseline and Capstone I--III controllers on one shared pipelining trajectory. Exceeding the cap only stops the affected controller. Copy the generated CSV/JSON files from \path{/aha/garnet/SIM_DIR} into the corresponding \path{data/capstone_<kernel>/} directory before the next run. Figure 12 additionally uses the documented \texttt{tensor3\_ttv} cap sweep. Rerun the Quick Start plotting command after exporting the regenerated data.

\subsection{Evaluation and expected results}

A successful Quick Start prints both:
{\footnotesize
\begin{verbatim}
REFERENCE VALIDATION: PASS -- bundled synthetic
 power-model data matches the reference signature.
REFERENCE VALIDATION: PASS -- bundled controller
 data matches the reference signature.
\end{verbatim}
}

The synthetic power-model workflow evaluates eight training and three held-out workloads and reports 7.54\% Capstone MAPE and 0.85\% oracle MAPE. The controller workflow detects eight kernel directories and reproduces Figures 5, 9, and 11--13 and Tables VII--IX from the released data.
The controller results should show Capstone I as the most conservative, Capstone II improving utilization, and Capstone III generally achieving the highest normalized frequency and lowest cap slack. Figure 12 should show nondecreasing frequency as the cap increases, while Table VIII should show complete bounds being more conservative than fit-only bounds.
The synthetic Figures 5, 9, and 10 are functional demonstrations rather than numerical reproductions because the original \ptpx{} hierarchy reports and row-level power values cannot be released due to NDA restrictions.

\subsection{Experiment customization}

Users may select a different kernel by changing the application-graph path in \path{/aha/cascade_demo.sh}. The power constraint is set with \texttt{CAPSTONE\_POWER\_CAP\_MW}. Controller modes, Capstone III bounds, and frequency/II scaling are configured through \\
\texttt{CAPSTONE\_RUN\_ALL\_MODES}, \\
\texttt{CAPSTONE\_BOUNDS\_JSON}, \\
\texttt{CAPSTONE\_REQUIRE\_TABLE8\_BOUNDS}, \\
\texttt{CAPSTONE\_USE\_FREQ\_II\_SCALING}, \\
\texttt{CAPSTONE\_FREQ\_REF\_MHZ}, and \\
\texttt{CAPSTONE\_II\_SCALE\_BY\_FREQ}.

The released runs use the kernel-specific
\texttt{CAPSTONE\_II\_ANCHOR\_Q\_MW} and
\texttt{CAPSTONE\_II\_SPEC\_Q\_MW} engineering margins listed in the README.
These values override \texttt{CAPSTONE\_II\_CALIBRATION\_JSON}, which should
remain unset unless both explicit margins are removed.

Users may also change the synthetic data seed, modify coefficients or bounds under \path{coeffs/}, or edit the Figure 12 cap grid in \path{plot/generate_figure12_sweep_data.py}. Omit \texttt{--validate-reference} when using modified data.

\subsection{Notes}

Figures 5, 9, and 10 under \path{figures/generated_power_model/} are synthetic functional demonstrations rather than numerical reproductions. The Figure 9 oracle fits the same model abstraction using all synthetic samples, including held-out samples, and therefore serves only as an optimistic reference.

The Baseline uses the guardband controller with an effectively infinite cap and consequently never stops the shared trajectory. Capstone I applies the paper's \(\gamma_{\mathrm{anc}}=0.45\) and \(\gamma_{\mathrm{spec}}=0.30\) guardbands. Capstone III loads its event and leakage bounds from \path{capstone_iii_bounds.json}. Table VIII evaluates the fit-only, doubled-fit, activity, PVT, and full OOD bound constructions, then repeats full-bound selection after pruning to \(K=8\) and \(K=4\).

The public Capstone II runs use the engineering margins \(q_{\mathrm{anc}}=0.064C\) and \(q_{\mathrm{spec}}=0.040C\), with \(\rho(f)=\max(1,f/100\,\mathrm{MHz})\), to demonstrate controller stopping and selection. The available calibration set is too small to support a finite-sample conformal guarantee, so these runs should not be interpreted as a new conformal-coverage result.

Table IX combines bundled Capstone results with values reported by prior work.

\subsection{Methodology}

Artifact reviewing and badging follow the ACM guidelines
at \url{https://www.acm.org/publications/policies/artifact-review-badging}
and the artifact-evaluation methodology at \url{https://cTuning.org/ae}.

\clearpage

\end{document}